\documentclass[default]{aastex631}
\usepackage{multirow}

\shorttitle{Photometric identification of unresolved binaries in open clusters}
\shortauthors{Mikhnevich et al.}

\graphicspath{{./}{figures/}}

\begin{document}

\title{Photometric Identification of Unresolved Binary Stars in Nearby Open Star Clusters.}

\correspondingauthor{Giovanni Carraro}
\email{giovanni.carraro@unipd.it}

\author[0000-0001-8514-3366]{Varvara O. Mikhnevich}
\affiliation{Ural Federal University,
19 Mira Street, 620002 Ekaterinburg, Russia}

\author[0000-0002-7504-0950]{Anastasiia Plotnikova}
\affiliation{Lund University, 
Box 188, 221 00 Lund, Sweden}

\author[0000-0002-0155-9434]{Giovanni Carraro}
\affiliation{Dipartimento di Fisica e Astronomia, Universita'  di Padova,
Vicolo Osservatorio 3, I35122, Padova, Italy} 

\author[0000-0001-8669-803X]{Anton F. Seleznev}
\affiliation{Ural Federal University,
19 Mira Street, 620002 Ekaterinburg, Russia}


\begin{abstract}

This paper introduces a new method to search for unresolved binary stars in open star clusters. 
The work aims at improving the approach introduced previously, which employs the (H-W2)-W1 versus W2-(BP-K) photometric diagram. 
This diagram, in tandem with the Gaia Color Magnitude Diagram (CMD) and using theoretical isochrones as reference sequences, is used to estimate the binary star fraction and the distribution of the component mass ratio $q$ in eight nearby open star clusters, including Pleiades, Alpha Per, and Praesepe, which we investigated in previous studies. 
In this study, to alleviate the uncertainties associated with the use of theoretical isochrones, we propose an empirical isochrones approach. 
We show that this is an effective approach to exploring a wider primary-mass interval, in particular for the region of low-mass sources.

Box-and-whisker plots are used to present the distribution of the component mass ratio $q$.
The mode of distribution turns out to be in the range $0.43-0.83$ and $0.38-0.63$ for Gaia and infrared-visible photometry, respectively. 
In addition, we update the algorithm to obtain the binary fraction, whose estimate lies in the range $0.16 - 0.36$ and $0.21 - 0.44$, depending on the adopted method, and show
that in previous studies the binary fraction was overestimated. 
 We do not find evidence that the variable spatial resolution of the employed catalogs (Gaia, 2MASS, and WISE) affects the precision of the binary fraction estimate.

\end{abstract}


\section{Introduction} \label{sec:intro}
Binary stars play a crucial role in the dynamical evolution of stellar populations in star clusters, both open and globular. In fact, mass transfer between stars in a binary system significantly alters their evolutionary paths, leading to the formation of exotic objects such as blue straggler stars \citep{Carraro2011MNRAS,Jadhav2021MNRAS,Harvey2024RNAAS} and cataclysmic variables \citep{Liebert1994AJ,Carraro2017ApJL}, routinely found in star clusters.
Open star clusters in particular are excellent laboratories for studying single- and binary evolution, as their stars are of similar age and initial composition. Among their members, most intermediate- and low-mass member stars are ideal tracers for investigating the properties of the cluster, such as binaries fraction, mass loss, and mass segregation \citep{Motherway2024ApJ,Zwicker2024ApJ,Angelo2023MNRAS,Angelo2025arXiv}. However, some observations of open clusters have revealed complexities that challenge the simple stellar population (SSP) paradigm, such as the presence of unresolved binaries and extended main sequences (eMSs) \citep{Spangler2025RNAAS,Chengyuan2024arXiv,Cordoni2023A&A}. Unresolved binary (UB) or multiple systems rise when two or more stars appear as a single point of light due to limitations in observational resolution. Unresolved systems influence the derivation of the initial mass function of stellar systems, thus providing insights into star formation theories \citep{Kroupa&Jerabkova2018,Kroupa2024arXiv}.

The presence of a significant number of unresolved binary systems in open star clusters was first suggested in the early 20th century \citep{Haffner&Heckmann1937}. 
Since then, studies have shown that the fraction of unresolved binary stars in open clusters typically exceeds 30\% \citep{Bonifazi+1990,Khalaj&Baumgardt2013,Sheikhi+2016,Sarro+2014}.
In addition, the distribution of the mass ratio $q=M_{sec}/M_{pri}$ (where $M_{pri}$ is the mass of the primary component and $M_{sec}$ is the mass of the secondary component) has been a topic of debate.

The flat distribution (all ratios have the same probability) has been the most commonly used. However, \cite{Kouwenhoven+2009} first proposed a power law for both field stars and clusters, and later a Gaussian distribution for star clusters \cite{Kouwenhoven+2009}, in agreement with \cite{Duquennoy&Mayor1991}. The presence of a peak near $q\gtrsim 0.95$ was found on the basis of spectroscopy \citep{Fisher2005MNRAS,Maxted+2008} and, more recently, using Gaia data \citep{El-Badry+2019,Alexander2025MNRAS,Childs2025arXiv}.  This peak was not present in the sample of \cite{Duquennoy&Mayor1991}. 
Finally, \cite{Patience+2002} noted that the distribution of $q$ varies for different mass intervals. 
The ongoing debate emphasizes the need for more comprehensive and accurate methods to detect and analyze these systems.

Identifying unresolved binaries in open clusters requires careful analysis. Various methods are utilized, such as photometric,  spectroscopic, or speckle interferometric observations.

\cite{BardalezGagliuffi+2014} proposed spectroscopic methods for identifying unresolved binaries among very low-mass stars and brown dwarfs. The periodic motion of stars in binary systems can be uncovered through radial velocity measurements. This process generally involves the use of medium to large telescopes and extensive monitoring over time to detect radial velocity variations. 
Speckle interferometry provides high angular resolution, allowing the detection of binaries with sub-arcsecond separation \citep{Chulkov2025AJ}. It can efficiently resolve binary stars in a relatively short amount of observing time compared to other methods like spectroscopy, but it is still influenced by seeing conditions, which can limit the achievable resolution and sensitivity. The effectiveness of the method can be reduced when there is a significant brightness difference between the components of a binary system, making it challenging to resolve those with a mass ratio less than $q=0.5$.

Photometric observations are relatively efficient as they can cover a large number of stars simultaneously, making them suitable for surveying entire clusters.
Identifying photometric binaries requires only a single observation epoch and does not have bias with respect to binary orbital period or inclination, making it a robust method for identifying a wide range of binary systems.
High-precision photometric data from surveys like Gaia, 2MASS and others can be used to recognize stars that are overluminous for their color, suggesting the presence of an unresolved companion. For example, \cite{Malkov+2010,Malkov+2011} suggested the use of Gaia photometric bands, which complement them with ultraviolet, but deep enough datasets in the ultraviolet range are currently unavailable.

Photometry and color-magnitude diagrams (CMDs) are essential tools for identifying and studying unresolved binary stars in open clusters. The presence of undetected binaries leads to a broadening of the main sequence (MS), which appears to be wider than what one expects solely from photometric errors. This occurs because the combined light from two stars in a binary system is brighter than that of a single star of the same color, causing the binary to appear above the main sequence. This broadening makes clusters appear older than they actually are \citep{Spangler2025RNAAS} and challenges the traditional assumption that stars within a cluster are coeval and share the same chemical composition \citep{Chengyuan2024arXiv}.

Several hypotheses have been proposed to explain the broadened MSs. Initially, age spread was proposed as an explanation, suggesting that clusters might have prolonged star formation histories. However, age differences alone may not fully explain the observed broadening \citep{Cordoni2023A&A}, such as the difference in chemical abundance. For example, the MS broadening of M~38 can be explained by differential reddening and unresolved binaries, without the need for a chemical abundance spread \citep{Griggio2023MNRAS}. Stellar rotation, particularly rapid rotation, has emerged as a plausible explanation. Rapidly rotating stars can exhibit lower effective temperatures and luminosities, causing them to appear redder and fainter than slowly rotating stars \citep{Chengyuan2024arXiv,Muratore2024A&A}. In the same vein, \cite{He2022ApJ} exclude the possibility that the MS split is caused by photometric errors or differential reddening. They found instead that slowly rotating stars on the red border of the MS may hide a binary companion.
 
Some clusters exhibit a bimodal distribution of rotational velocities, with a population of slow rotators and another one of fast rotators. The origin of this distribution is still debated, with hypotheses including tidal locking in binary systems or differences in the pre-main sequence rotation rates \citep{Chengyuan2024arXiv}. \cite{Wang2023ApJ} investigate whether the long-term dynamical evolution of star clusters can produce the observed split in the MSs of young massive clusters, using high-performance N-body simulations. They concluded that while tidally locked binaries and blue straggler stars can contribute to the split MS phenomenon, their numbers in the simulation are insufficient to fully explain the observations, particularly in NGC 1856. With respect to unresolved binary stars, they have the potential to significantly affect the observed properties of star clusters. Therefore, investigating unresolved systems seems to be important, considering all relevant information.

Several recent studies have investigated binaries in open clusters using color-magnitude diagrams with Gaia photometric data. Detecting binaries with a low mass ratio ($q<0.5$) remains difficult with optical data alone, despite the high-precision Gaia photometry. However, it is crucial to consider such systems. They are more sensitive to dynamic interactions because of the smaller binding energies. That makes them very important for studying dynamic effects. The stellar mass distribution cannot be fully understood without the presence of binaries with low mass ratios. Also, proper estimation of binaries fractions provides more accurate constraints on the theories of star formation. 

Multi-band photometric data are essential in methods based on photometry. For example, analyses based only on optical photometry usually adopt mass ratio limits of about $q=0.5 - 0.7$ \citep{Cordoni2023A&A,Donada2023A&A,Jiang2024ApJ}. When such binaries are better identified using a combination of optical and infrared photometry, this can extend the limit of the mass ratio to $q=0.2 - 0.4$ \citep{Malofeeva+2022,Malofeeva+2023,Childs2024ApJ,Liu2025AJ}. 
However, some binary systems with $q<0.2 - 0.3$ remain complicated to identify with certainty.\\

The present study is organized as follows.
Section~\ref{sec:Review} is devoted to a review of the literature on the subject.
In Section~\ref{sec:Samples} we comprehensively discuss the membership of the clusters.
In Section~\ref{sec:Empirical_isochrones} we describe the algorithm for building empirical isochrones.
Section~\ref{sec:Main_results} presents our results on the binaries fraction and the mass ratio distribution.
Section~\ref{sec:Chance_matches}, then, is dedicated to the investigation of the chance matches in probable members samples of open star clusters.
In Section~\ref{sec:Method_reliability} we perform a numerical test to evaluate the applicability of our method for unresolved binary identification.
Finally, Section~\ref{sec:Discussion} performs a thorough discussion of our findings. Section~\ref{sec:Conclusion} wraps main results.

\section{Methodology} \label{sec:Review}
\citet{Malofeeva+2023} investigated the population of unresolved binary and multiple star systems in four open star clusters: Pleiades, Alpha Persei, Praesepe, and NGC 1039. 
The study used a multiband photometric combination, (H-W2)-W1 vs W2-(BP-K), to identify and analyze these systems --- two-index diagram (TID), introduced by \cite{Malofeeva+2022}, which combines optical and infrared photometry to enhance sensitivity to low-mass companions. The selection of these photometric indices was made by examining all possible combinations of two and three bands.
Theoretical isochrones and lines with constant $q$ values (isolines) were used to model the distribution of binary and multiple systems.

The rationale for combining optical and infrared data lies in the spectral energy distribution (SED) of multiple systems. Low-mass companions (e.g., M dwarfs or brown dwarfs) emit most of their flux in the near-infrared (H, K) and mid-infrared (W1, W2). While such stars contribute little excess brightness in the visible (G band), they produce significant excess in the H, W1, and W2 bands. By constructing composite colors that amplify these differential flux contributions, the TID stretches the photometric separation between single stars and binaries. It makes even low-luminosity companions detectable through small but systematic shifts away from the single-star sequence, when the region occupied by small $q$ binary systems collapses when using only optical colors.
The foundation of the method has been refined in \cite{Malofeeva+2023} to improve accuracy and error estimation.
The key improvements over the previous work were the following:
\begin{itemize}
    \item Automatic star counting: the procedure for star counts has been automated to reduce the human factor.
    \item Bootstrapping: conducted 100–300 Monte Carlo simulations per cluster to estimate photometric uncertainties.
    \item Modeling of multiple systems: the study includes modeling of triple and quadruple systems to better constrain the regions they occupy in the photometric diagram.
\end{itemize}

A formalism for the binary fraction parameter $\alpha$ is commonly defined as:

\begin{equation}
\label{alpha}
\alpha=\frac{N_{binaries}+N_{triples}+N_{quadruples}}{N_{singles}+N_{binaries}+N_{triples}+N_{quadruples}}\; .
\end{equation}

The total binary fraction $\alpha$ in the four studied open clusters ranges from $0.45\pm 0.03$ to $0.73\pm 0.03$. 
These values exceed earlier estimates. 
The discrepancy was attributed to the improved sensitivity in detecting low-mass-ratio systems.
Systems with multiplicity greater than two constitute $6- 9\%$ multiple systems in clusters, with triples dominating over quadruples.
The distribution of the component mass ratio $q$ is well fit with a Gaussian.
The mode of distribution ranges from $0.22\pm 0.04$ to $0.52\pm 0.01$, and the dispersion ranges from $0.1\pm 0.02$ to $0.35\pm 0.07$.

However, the methodology has notable limitations.
Firstly, theoretical isochrones \citep{Bressan+2012} with \cite{Dias+2021} fundamental parameters exhibit disagreement with observed main sequences (MS) in the interval of sources with $M_{pri}<0.5M_{\odot}$ and do not perfectly match the remaining one. 
That leads to limitations in the mass range explored ($0.5 - 1.8M_{\odot}$) and uncertainties in the binary fraction due to misplaced isolines. 
We propose a solution to this problem in section~\ref{sec:Empirical_isochrones}. 

Secondly, photometric errors and the presence of field stars in samples \citep{Lodieu+2019,Danilov&Seleznev2020,Nikiforova+2020,Cantat-Gaudin+2020} can also affect the observed binary fraction. 

Concerning the low-mass region ($<0.5M_{\odot}$), it should be noted that there are many sources that are located below and to the left of the main sequence on the color-magnitude diagram. 
That might appear to be unresolved binary systems consisting of a main sequence star and a white dwarf. 
In \citet{Mikhnevich&Seleznev} we modeled these systems to understand their photometric properties and how they appear on the color-magnitude diagram.

\section{Cluster's members samples}\label{sec:Samples}
With the aim of deriving the ratio of unresolved binaries, we start by assessing the membership of the clusters under analysis.
In this study, we extract stars from the \citet{Hunt&Reffert2024} updated large homogeneous catalog of the Galaxy open star clusters. 
This catalog is based on the latest Gaia data release DR3 \citep{GaiaDR3}. 
\begin{figure}[h!]
    \centering
    \includegraphics[width=0.7\linewidth, keepaspectratio=1]{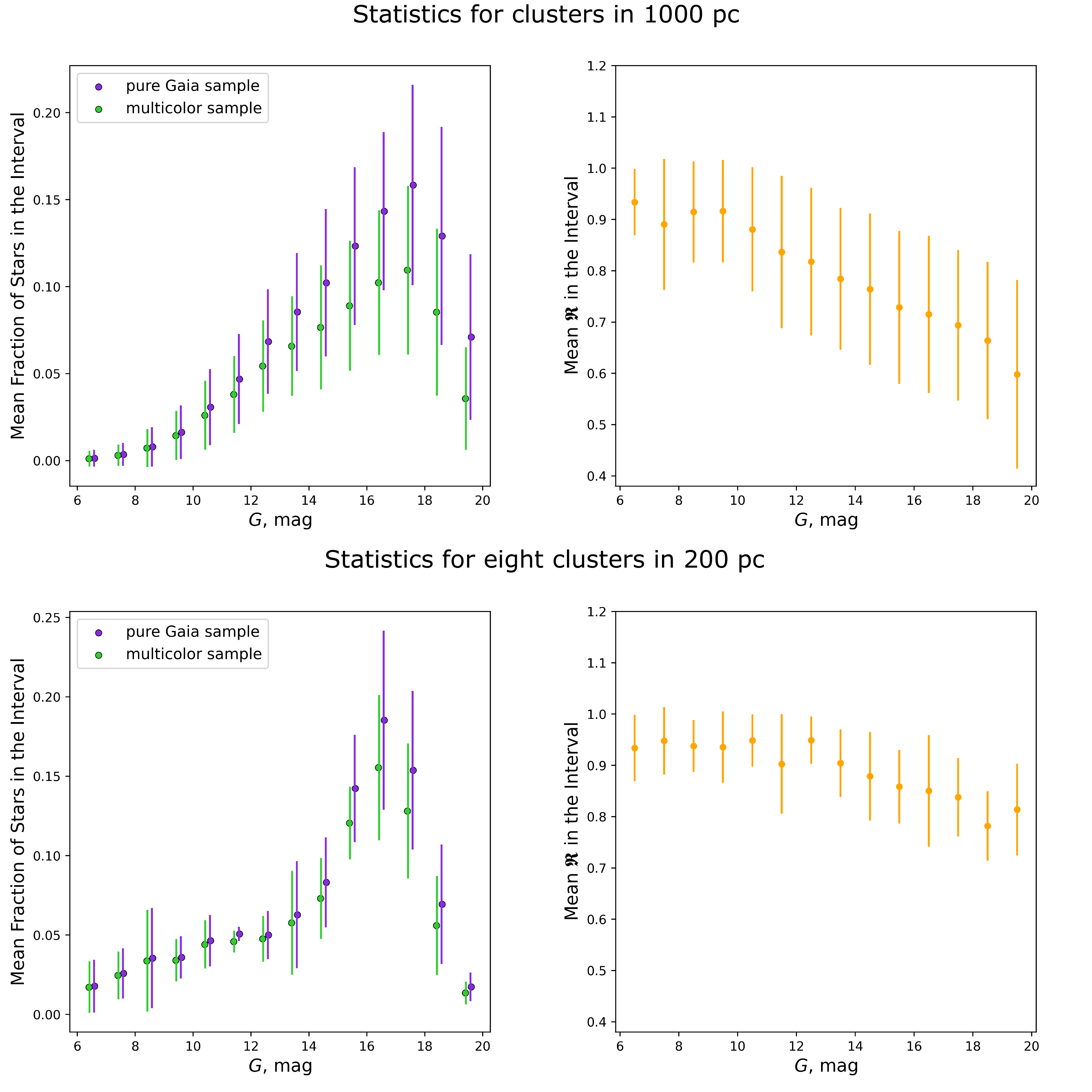}
    \caption{The statistics for member's samples. 
    {\bf Left:} Mean fraction of stars in the samples, binned into 1-magnitude-wide intervals of G magnitudes, with error bars representing uncertainties for both the pure Gaia (purple dots) and multicolor (green dots) samples. 
    The fraction is obtained relative to Gaia samples size individually for each cluster. 
    {\bf Right:} The mean ratio $\mathfrak{R}=N_{mc}/N_{Gaia}$ of number of stars in 1-magnitude-wide intervals of G magnitudes in multicolor sample $N_{mc}$ and the same in Gaia sample $N_{Gaia}$.
    In the top line samples of clusters within 1 kpc are accounted. In the bottom line eight clusters within 200 pc are considered: IC~2391, IC~2602, Melotte~111, Melotte~20, Melotte~22, Melotte~25, NGC~2451A,  and NGC~2632.}
    \label{fig:mean_frac_in_int}
\end{figure}
The filters and their effective wavelength ranges are as follows: G (Broadband visible light) covers $\sim330-1050$ nm with central $\lambda \approx 640$ nm; BP (Blue Photometer) --- $\sim330-680$ nm; RP (Red Photometer) --- $\sim640-1050$ nm. 
We consider the membership lists complete up to the 20 magnitude G-band and limit ourselves to including members with probability greater than 50\%. 
To be able to use the multicolor approach, we cross-matched these lists with the 2MASS and WISE catalogs \citep{Skrutskie+2006,Wright+2010,Mainzer2011ApJ} for mid-infrared bands W1 and W2, corresponding to $\sim3.4\;\mu$m, $\sim4.6\; \mu$m, respectively, and near-infrared bands H and K, corresponding to $\sim1.65\; \mu$m and $\sim2.17\; \mu$m, respectively.

In order to maximize the number of matches in the three catalogs, we implement the xMatch queries. 
The xMatch service is a source cross-identifying tool of the Astroquery coordinated package of Astropy \citep{astroquery,astropy:2013,astropy:2018,astropy:2022}. 
The service identifies stellar matches by comparing source coordinates to a common epoch. 
Inevitably, the matching has to deal with the different spatial resolutions of the different catalogs. 
In fact, Gaia has a resolution of up to 1.5 arcsec, while 2MASS and WISE have resolutions of $\sim4$ and $\sim6.4$ arcsec, respectively \citep{gaiaedr3_doc,2mass_doc,Wright+2010}. 
As a result, discrepancies arise, since a single star recorded in 2MASS or WISE may correspond to multiple sources in Gaia, leading to potential false matches. 
We investigate how false matches affect the results for the binary fraction and the mass ratio distribution in section \ref{sec:Chance_matches}.

In view of the above issue, the samples is also reduced in terms of number of sources. 
This is shown in the left panels of Figure \ref{fig:mean_frac_in_int}. 
The average fraction of stars in the sample, binned into 1-magnitude-wide intervals of G magnitudes, exhibits a notable difference in the $16^m-20^m$ region. 
Generally, the mean relation $\mathfrak{R}=N_{mc}/N_{Gaia}$ between the number of stars in 1-magnitude-wide intervals of G magnitudes in the multicolor sample ($N_{mc}$) and in the Gaia sample ($N_{Gaia}$) tends to decrease as the magnitudes become fainter. This is shown in Figure \ref{fig:mean_frac_in_int}, top right panel.
These facts could limit the approach when considering distant clusters.
To ensure a comprehensive analysis and to take advantage of all the potential of the multicolor approach, we also applied the same methodology to pure Gaia samples and examined the differences. For this purpose, we have selected appropriate clusters.

We consider eight open star clusters, all located within 200 parsecs: IC~2391, IC~2602, Melotte~111 (Coma), Melotte~20 (Alpha~Per)\footnote{While some studies classify Alpha Persei (Melotte 20) as a moving group (e.g., \cite{Wright2020NewAR}), it is treated as an open cluster in the \citet{Hunt&Reffert2024} catalog --- our primary source for membership selection --- and shares kinematic and spatial characteristics with other clusters in our sample. For consistency, we follow their classification.}, Melotte~22 (Pleiades), Melotte~25 (Hyades), NGC~2451A, and NGC~2632 (Praesepe).
The samples of these clusters exhibit discrepancies of approximately 5\% or even less in their contents in the same magnitude region (Fig. \ref{fig:mean_frac_in_int}, bottom line, left panel). In Figure \ref{fig:mean_frac_in_int} in the bottom line, right panel, an almost flat distribution of the relation is evident.
Thus, we can safely investigate these star clusters with minimal risk of incorporating uncertainties due to sample disparities.

In addition, the mean relation $\mathfrak{R}$ in 1-magnitude-wide intervals of G magnitudes tends to 1.0 for clusters located outside of the Galactic disk, and does not have any notable statistical dependence on extinction $A_V$ or distance to the cluster. 
\begin{figure}[h!]
    \centering
    \includegraphics[width=\linewidth, keepaspectratio=1]{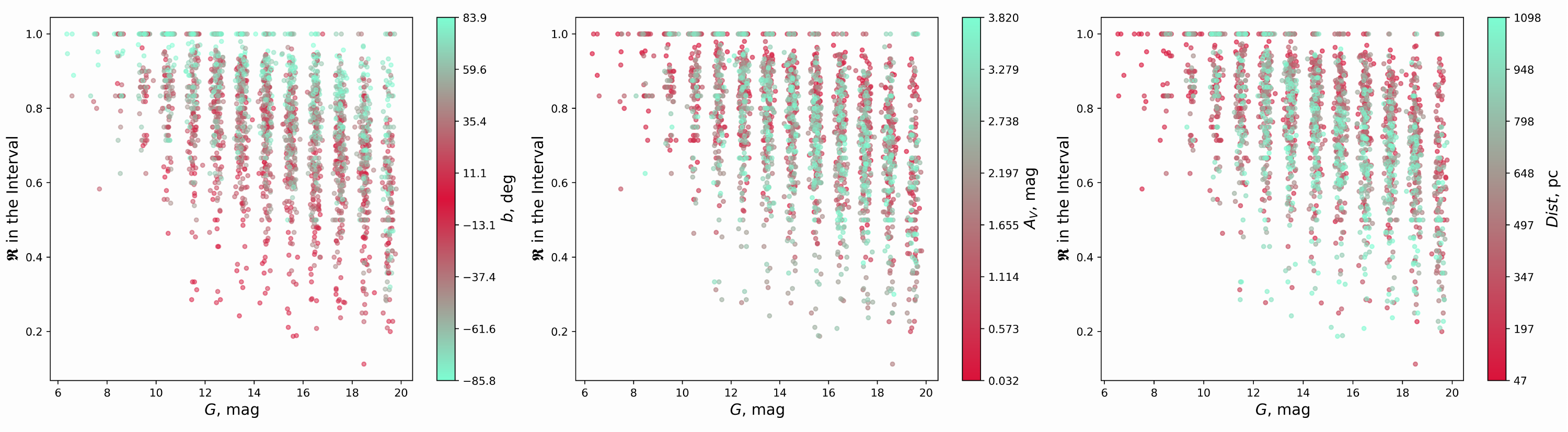}
    \caption{Mean relation $\mathfrak{R}=N_{mc}/N_{Gaia}$ between number of stars in 1-magnitude-wide intervals of G magnitudes in multicolor sample $N_{mc}$ and in Gaia sample $N_{Gaia}$.
    Samples of about 300 clusters in 1000 pc are accounted. 
    The color-bars represent variations in latitude ($b$), extinction ($A_V$) and distance ($Dist$) for each cluster. 
    The dots are randomly spread around the $G_i$ magnitude of each interval.
    The fundamental star cluster parameters were taken from the catalog of \citet{Dias+2021}.}
    \label{fig:colorbar_relation_of_frac_in_int}
\end{figure}
This is illustrated in Figure \ref{fig:colorbar_relation_of_frac_in_int}. 
For visualization purposes, the dots are randomly spread around the magnitude $G_i$ of each interval with $\sigma=0.12$.

\section{Empirical isochrone construction} \label{sec:Empirical_isochrones}
As we mentioned earlier, theoretical isochrones, particularly the line representing single stars ($q~=~0$), do not coincide with the cluster main sequence across the entire range of stellar magnitudes. 
This discrepancy is more pronounced in the lower part of the diagrams\textbf{, where $M_{pri} < 0.5M_{\odot}$}, especially in the two-index diagram (TID). 
The discrepancy arises from uncertainties in model atmospheres, extinction laws, and unresolved binaries contaminating the training samples used to calibrate models (e.g., \citet{Bressan+2012,Danilov&Seleznev2020}). These mismatches compromise the accuracy of binary detection via photometric displacement, leading to a potential overestimation or underestimation of single and binary star counts.

To overcome this limitation, we adopt an empirical approach, constructing a data-driven representation of the single-star sequence for each cluster individually --- an empirical isochrone \citep{Mikhnevich2024maeu.conf}. Our method leverages the fact that, in a well-sampled cluster, single stars form a dense, continuous concentration along the MS in CMDs and TIDs, while unresolved binaries appear systematically offset toward brighter magnitudes and redder colors.

We identify the high-density concentration using the HDBSCAN\footnote{The HDBSCAN algorithm is implemented via the \texttt{hdbscan} Python package \citep{McInnes2017}, which is built on the theoretical framework described in \citet{McInnes2017arXiv}} (Hierarchical Density-Based Spatial Clustering of Applications with Noise). It is a robust, density-based multidimensional clustering algorithm that identifies clusters as regions of high density separated by regions of low density, while also detecting and excluding noise (outliers). The main reason for exploitation is its sensitivity to data with uneven density, which applies to the morphology of the cluster main sequence as we use photometry as input data. The core idea of HDBSCAN is to build a hierarchy of clusters across all possible density levels, then extract the most `persistent' clusters --- those that exist over the widest range of density scales. The highest-density contiguous structure identified corresponds to the single-star sequence, while stars above and to the right on CMD or to the left on TID --- classified as noise or minor clusters --- are candidate unresolved binaries.

The algorithm's behavior is mainly influenced by two parameters that have distinct roles: \texttt{min\_cluster\_size} and \texttt{min\_samples}. 

\texttt{min\_cluster\_size} sets the minimum number of stars required for a grouping to be considered a valid cluster during hierarchical condensation. Substructures smaller than this threshold are pruned and treated as either noise or part of a broader structure.

\texttt{min\_samples} defines the number $k$ of nearest neighbors for the considered point, used to estimate local density. Mathematically, it sets the core distance $\kappa(x_i)$ of a star $x_i$ as the distance to its $k$-th nearest neighbor: $\kappa(x_i)=dist(x_i, x_{i,k})$. This acts as a smoothing scale: larger values produce smoother density estimates and reduce sensitivity to small fluctuations or measurement errors.

The application of the algorithm to both diagrams requires careful consideration of the morphological features of the clusters. 
For the CMD in Gaia bands, the algorithm uses the entire cluster sequence; however, the TID is divided into higher-density and lower-density sections to better account for its morphological features.

The parameters \texttt{min\_cluster\_size} and \texttt{min\_samples} set in HDBSCAN are presented in Appendix~\ref{appendix:hdbscanPar}.
We consider only those sources for which HDBSCAN assigns a cluster membership probability larger than 50\%. 

\begin{figure}[h!]
    \centering
    \includegraphics[width=\linewidth]{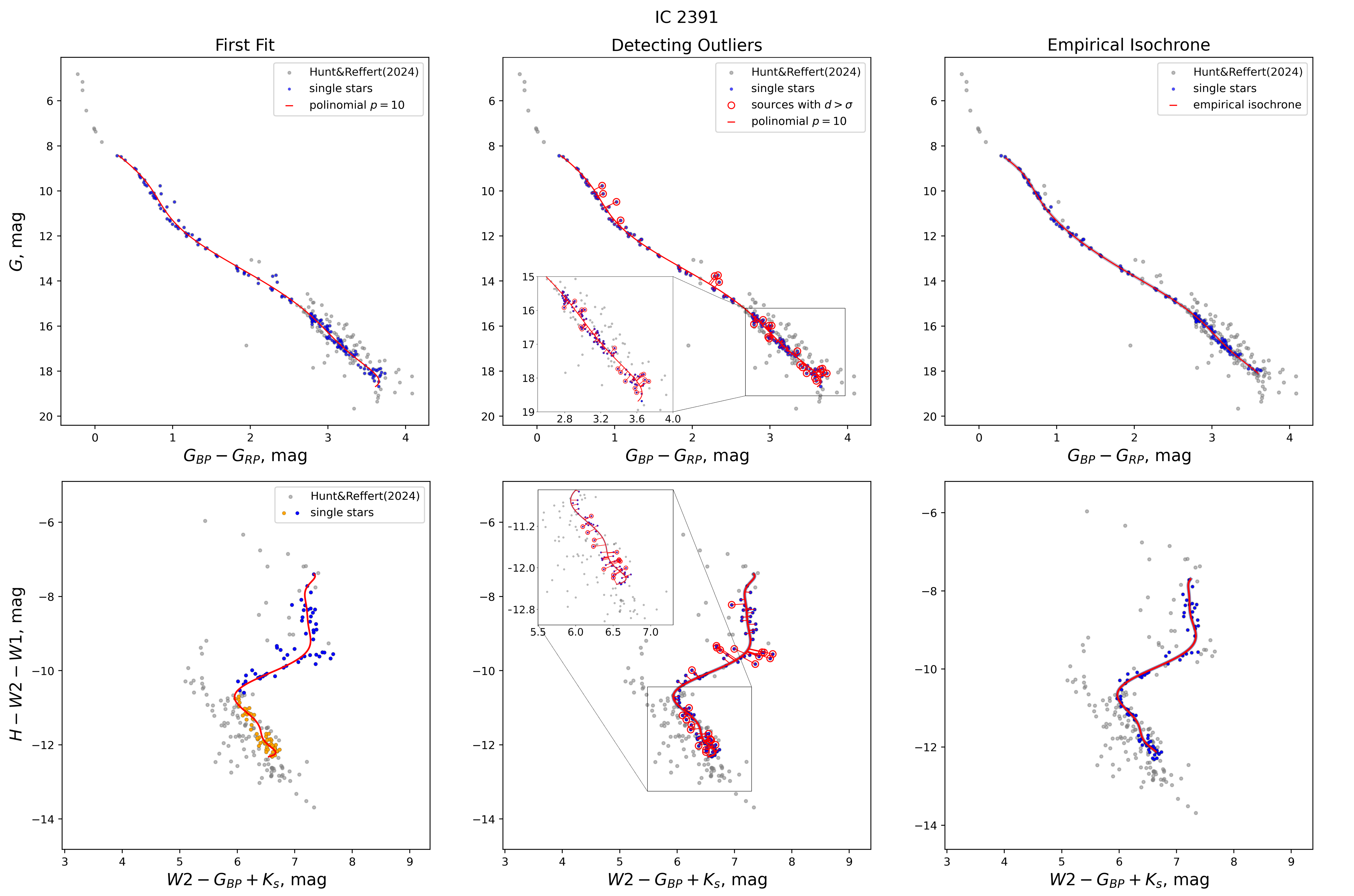}
    \caption{The three steps process to remove outliers in the CMD (top panels) and in the TID (bottom panels) for IC~2391. 
    Axes display apparent magnitudes. 
    The left panels show the initial approximation of the sample assigned by HDBSCAN, where TID main sequence is separated because of differential data density. 
    This initial fit is far from being optimal due to presence of outliers, which distort the overall alignment. 
    We remove outliers groups implementing the algorithm described in section~\ref{sec:Empirical_isochrones}. 
    Following this, we perform the second approximation and eliminate individual outliers, as shown in the central panels.
    The inserted frames illustrate the higher-density part of the cluster sequence in enlarged scale.
    We then perform the final approximation and consider it as the empirical isochrone for single stars. This is shown in the right panels.}
    \label{fig:emp_iso_alg}
\end{figure}

Nevertheless, the samples still contain concentrations of binary stars that shift the empirical isochrone away from the single star sequence. To exclude these concentrations, we follow this algorithm: 

\begin{itemize}
    \item Perform an initial approximation of the sample with a polynomial of $10^{th}$ degree and determine the deviation $d_{ij}$ of each $j$-source in each $i$-group of \texttt{hdbscan.labels} from the polynomial. 
    \item In each $i$-group, the mean deviation $\bar{d_i}$ of the source from the polynomial and the standard deviation $\sigma$ of this value are determined.
    \item  If the value of $\bar{d_i}$ for one of the groups differs from the average value $\bar{d}$ for all the groups by more than $1\sigma$, all elements of this group are considered outliers and excluded from the sample. 
\end{itemize}

However, this approach is not sensitive to individual sources.
For better filtering, we perform a second approximation of the updated sample and determine the deviation $d_{ij}$ again, from the new polynomial. 
This time, we estimate $\bar{d}$ and $\sigma$ separately for the higher and lower density parts of the cluster sequence (for the CMD, the boundary is approximately at G=15, and for the TID, at ((H-W2)-W1=-10.5, respectively). 
Finally, we exclude sources with deviations greater than $1\sigma$ from the mean values. This is presented in Fig. \ref{fig:emp_iso_alg} in the diagrams for IC~2391, providing clear representation due to the smallness of its sample.
We refer to the final approximation of the cleaned sample as an empirical isochrone of single stars (EIS) in the cluster. In Appendix~\ref{appendix:EIs} diagrams for all other clusters are presented.

\begin{figure}[h!]
    \centering
    \includegraphics[width=\linewidth]{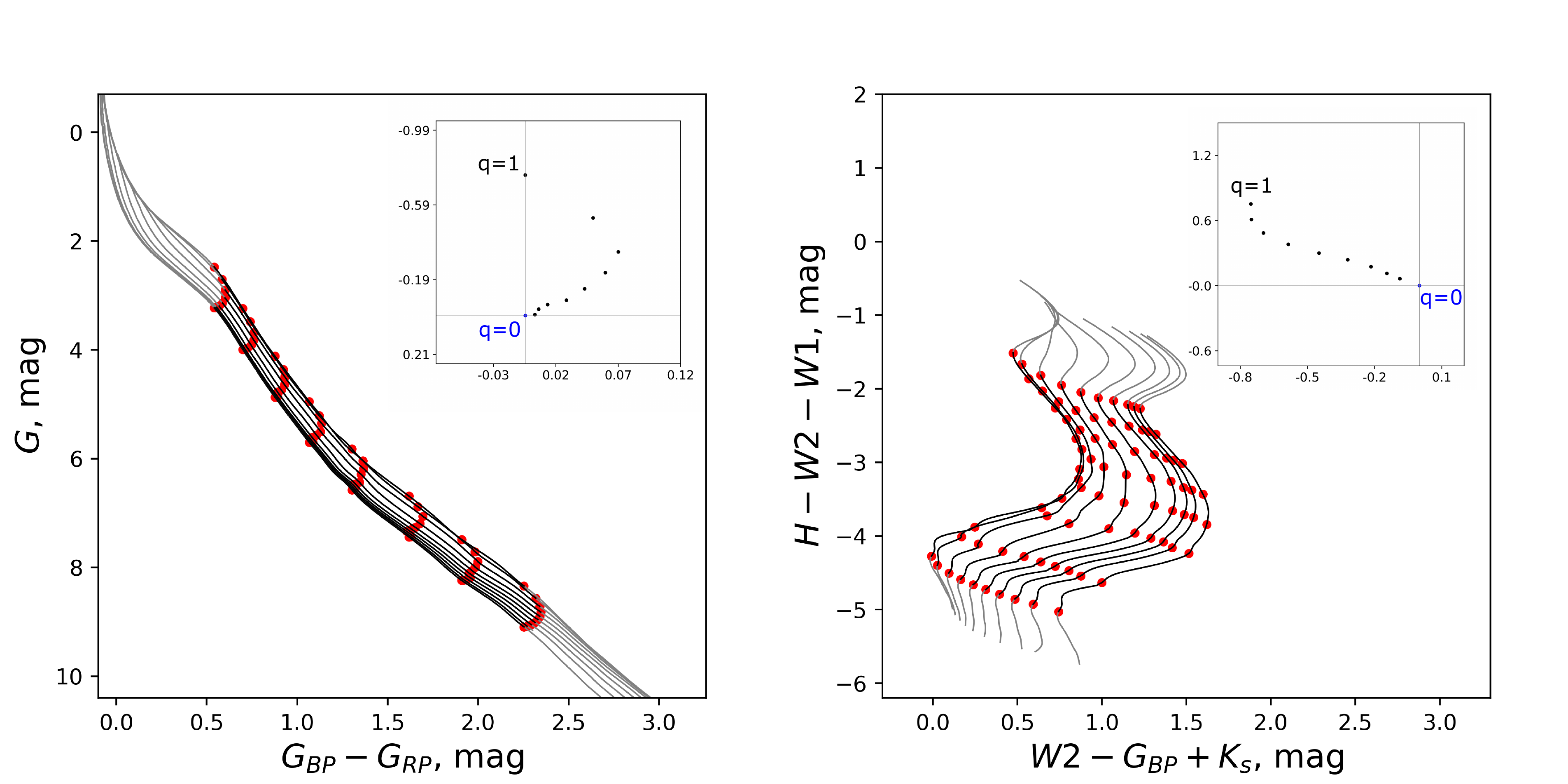}
    \caption{An illustration of our method to determine the shift values for the empirical isochrones of binary stars. 
    {\bf Left}: PARSEC theoretical isochrones for single stars and binary stars with different $q$ in CMD. {\bf Right}: the same for TID. Axes display absolute magnitudes. We use the red dots to define the average shift values for the empirical isochrones of binary stars. These average shift values are shown in inserted frames zoomed in relative coordinates. The crosshair indicates the position of the isochrone of single stars $q=0$, i.e. this is the starting point from which all shifts are measured. 
    The large shift ($\sim1$ mag) along the abscissa in TID reflects the high sensitivity of the composite color (H-W2)-W1 to infrared flux excess from unresolved companions. Because cool secondaries emit strongly in the mid-infrared (especially W2), even low-mass companions cause significant reddening in this index.}
    \label{fig:shifts_from_theor_iso}
\end{figure}

Empirical isochrones for unresolved binary systems (EIBs) with mass ratios $q$ ranging from 0.2 to 1.0 in increments of 0.1 are calculated by shifting the empirical isochrone for single stars according to the specific values determined for each cluster.
These values are the average distances between the theoretical \mbox{PARSEC} isochrone \citet{Bressan+2012} for single stars and the theoretical isochrones for binary stars with different $q$ in several zones of the MS (Fig.~\ref{fig:shifts_from_theor_iso}). 
We plot the theoretical isochrones for binary stars using the theoretical isochrone for single stars and a simple relationship for unresolved binary star magnitude derived from the definition of stellar magnitude:
\begin{equation}
    \begin{array}{c}
        m_{bin}=m_1-2.5 \log{\left( 1+\frac{1}{x} \right)}, \\
        \text{ where } x=10^{-0.4 (m_1-m_2)},
    \end{array}
\end{equation}
$m_{bin}$ is common magnitude of a binary system, $m_1$ and $m_2$ are magnitudes of the primary and the secondary components respectively. 
We extract the fundamental star cluster parameters for theoretical isochrones from the catalog of \cite{Dias+2021}. The mass interval used for the primary component starts with 0.5$M_{\odot}$, due to the limits of the theoretical isochrones, and ends with the maximum mass of the empirical isochrone constructed.

\section{Binary fraction and mass ratio distribution} \label{sec:Main_results}
\begin{figure}[h!]
    \centering
    \includegraphics[width=\linewidth]{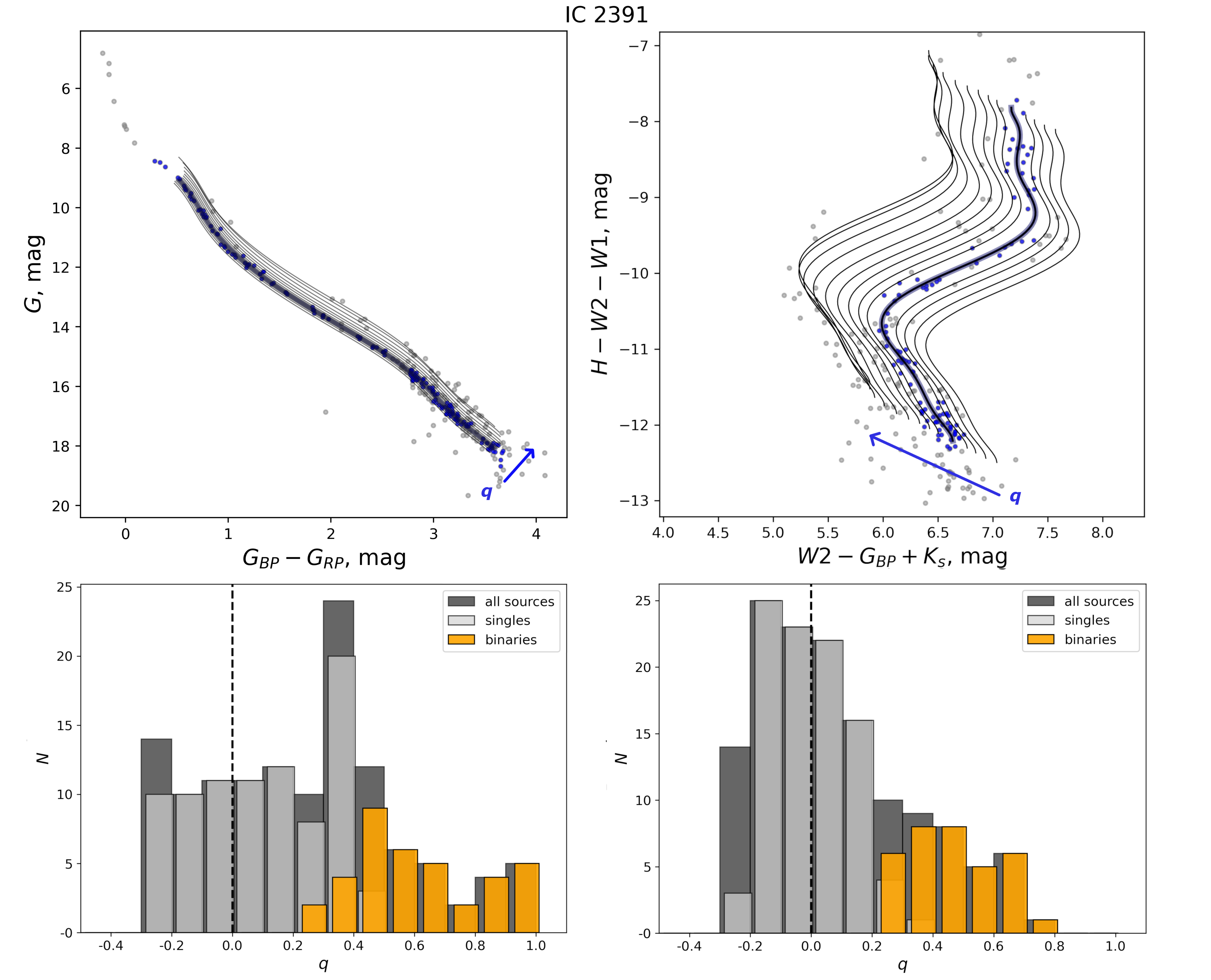}
    \caption{In the top line, CMD and TID are presented with empirical isochrones of binary stars with $q=-0.5, -0.4, -0.3, -0.2, 0.0, 0.2, 0.3, 0.4, 0.5, 0.6, 0.7, 0.8, 0.9, 1$ indicated by thin black lines.
    The blue arrows indicate the direction of increasing $q$-value.
    The thick dark-blue line represents the empirical isochrone of single stars.
    In the bottom panels, the mass ratio distributions of binaries, extracted from corresponding diagrams, are indicated in orange. 
    They result from the subtraction single-sources distribution (light-gray histogram) from all-data distribution (dark-gray histogram).}
    \label{fig:diag_for_counts}
\end{figure}

The CMD and TID for IC~2391 with the empirical isochrones are presented in the upper panels of Fig.~\ref{fig:diag_for_counts}. 
Counts are performed in the intervals between isochrones. 
We calculate the specific value of the parameter $q$ for each source by linear interpolation between two EIBs of $q_n$ and $q_{n+1}$ in the direction of the shift vector from the theoretical isochrones.
To consider the dispersion of stars on both sides of the empirical isochrone for single stars, we introduce the `negative' $q$ values down to $q=-0.5$. EIBs for these values are located to the left of and below the cluster sequence in CMD, and to the right in TID.
Then we present boxplots for a distribution of the estimated component mass ratio $q$. 
To estimate the distribution for binaries, we subtract the distribution for sources, assigned by HDBSCAN as single stars and situated nearer to EIS than EIB for q=0.1, from the distribution calculated for all data in the optimal mass interval, particularly $M=[0.15;1.69]M_{\odot}$ for IC~2391 (the bottom line of Fig.~\ref{fig:diag_for_counts}). 

\begin{figure}[h]
    \centering
    \includegraphics[width=0.9\linewidth]{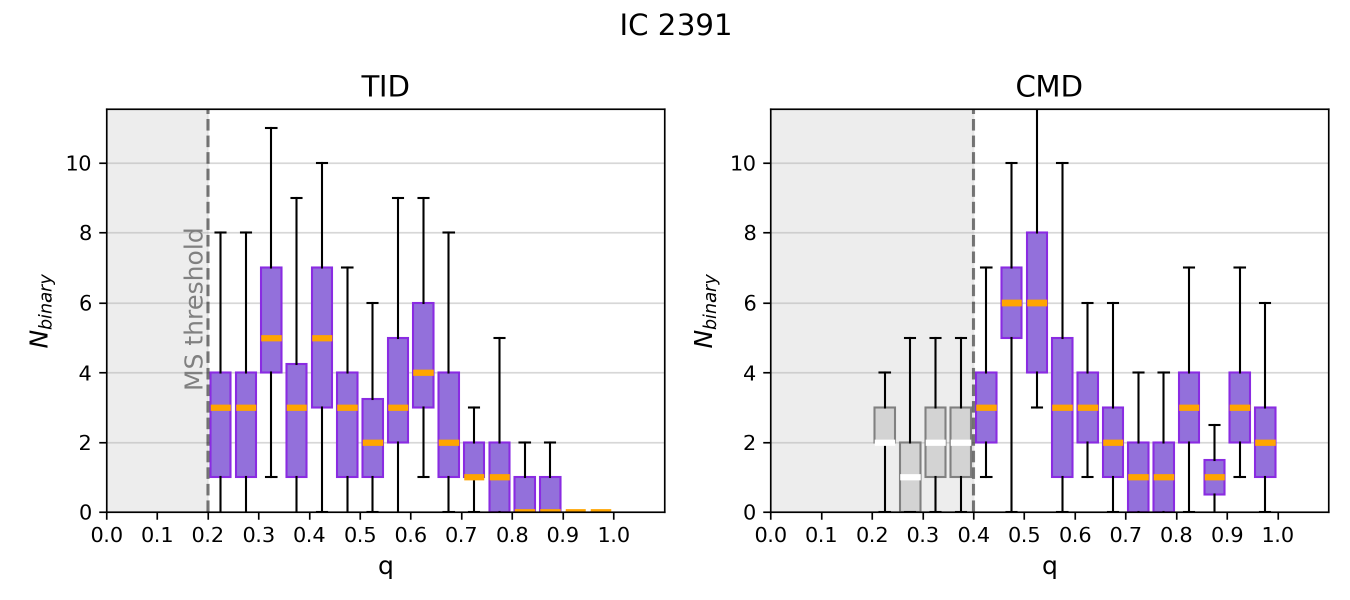}
    \caption{The box-and-whisker plots of mass ratio distributions of binary systems derived from CMD (right panel) and TID (left panel) for IC~2391. 
    Orange dashes display initial counts from observations (particularly in this representation), purple boxes are 25$^{th}$ and 75$^{th}$ quartiles, and the most extreme non-outlier data points (whiskers) per q-interval from 100 realizations are denoted in black.
    }
    \label{fig:var_distribution}
\end{figure}

To account for photometric errors, we went back to the initial step. 
Before assigning the single-star sequence with HDBSCAN, we vary the magnitudes of the sources according to their errors, as indicated in the catalogs. 
To this end, we use \texttt{ numepy.random.normal}, assuming a normal distribution of errors. 
Then, the entire algorithm, up to the estimation of the binaries fraction in the cluster, is run 100 times. 
In Figure~\ref{fig:var_distribution}, one can notice the purple intervals of possible $q$ values as the interquartile range (IQR), affected by photometric errors. 
The central orange markers (and, in addition, the white markers for CMD) do not represent medians from a sample of variations. Instead, they represent the direct output of the pipeline using non-variated photometric values from the real catalog data. 
This approach is chosen to preserve the specific, observed properties of the clusters rather than presenting purely stochastic results.

Looking at the distribution in the CMD in Fig.~\ref{fig:diag_for_counts} (lower panel) , a dip near q = 0 can be observed.
We argue that this is caused by the extremely close relative positions of the EIS and EIBs with q = [0.2;-0.2]. 

To explain this feature, we compare the Gaia photometric error and the spread of sources, assigned as singles, around EIS for each cluster. 
The result is illustrated in Fig.~\ref{appendix_fig:ms_err} in Appendix~\ref{appendix:ms_errors}.
Here, the distances between the EIBs are smaller than the photometric errors of the singles, whereas the spread of these sources around the EIS is usually larger. 
Therefore, singles go beyond the closest isochrones for unresolved binaries, and the low-mass ratio region is less abundant in sources when compared to other intervals.

This limitation is not present when working with the TID. In Figure~\ref{appendix_fig:ms_err}, one can notice that the scales on the axes are different, the errors are larger for multicolor photometry, but the distance to the EIBs allows for greater confidence.
This limits the range of explorable values of the q parameter using the method used in this work. For CMD the mean limit is $q_{CMD}>0.4$, for TID it is $q_{TID}>0.2$ (Fig.~\ref{fig:comparative_diagram}, \ref{appendix_fig:ms_err}). 
In the distribution from TID, we observe a tendency to have large amounts of the low-mass-ratio sources. It is notable because of the more distinguishable EIBs in TID.

In \textbf{the majority of the} clusters, the distribution of binary stars shows a maximum. 
The mode of distribution $q_{mode}$ using CMD is between 0.43 and 0.83, while using TID the results vary from 0.38 to 0.63.
The graphics for each cluster are presented in Appendix~\ref{appendix:qDistributions}.

The fraction of binary stars in the cluster $\alpha$ is extracted from the binned distribution. 
Depending on CMD or TID, $\alpha$ fluctuates between 0.16 and 0.36 or between 0.21 and 0.44.
The results are presented in Table~\ref{tab:binary_results}. It is important to note that multiple systems, which were located to the right of q = 1 in the CMD and to the left in the TID, are not taken into account.

The obvious advantage of TID is that it allows one to identify a larger fraction of unresolved binaries including very low q-value sources.
These sources might constitute the majority of unresolved binaries, as can be inferred from the $q_{mode}$ values for clusters.  
In the case of $M_{pri}\approx 0.15-0.20 M_{\odot}$ and $q\lesssim0.35-0.53$, the secondary component is a brown dwarf that would be extremely difficult to identify as a single object. 
Thus, it is essential to account for low-q sources.

\begin{deluxetable*}{lccccccccccc}
    \tabletypesize{\scriptsize}  
    \tablecolumns{12}
    \tablewidth{\textwidth}
    \tablecaption{Binary Population Parameters for Eight Nearby Open Clusters: Results from Full and Truncated Samples. 
    Columns (4)--(7) present binary fraction ($\alpha$) with 25$^{th}$ and 75$^{th}$ quartiles and mode of the mass ratio distribution ($q_{\text{mode}}$) derived from the full sample. 
    Columns (8)--(11) show the same parameters for the truncated sample (excluding likely false matches). 
    Column (12) gives the relative loss of sources due to exclusion of close neighbors, defined as $1 - N_{\text{trunc}}/N_{\text{full}}$.
    \label{tab:binary_results}  }
    \tablehead{
    \multirow{3}{0.05\textwidth}{Cluster} & 
    \multirow{3}{0.05\textwidth}{$M_{\min}$ ($M_\odot$)} & 
    \multirow{3}{0.05\textwidth}{$M_{\max}$ ($M_\odot$)} &
    \multicolumn{8}{c}{Binary Parameters} &
    \multirow{2}{0.05\textwidth}{\shortstack{Relative \\ Loss}} \\
    \cline{4-11}
    & & & \multicolumn{4}{c}{Full Sample} & \multicolumn{4}{c}{Truncated Sample} & \\
    \cline{4-11}
    & & & \multicolumn{2}{c}{TID} & \multicolumn{2}{c}{CMD} & \multicolumn{2}{c}{TID} & \multicolumn{2}{c}{CMD} & \colhead{$1 - \frac{N_{\text{trunc}}}{N_{\text{full}}}$} \\
    \cline{4-11}
    & & & $\alpha$ & $q_{\text{mode}}$ & $\alpha$ & $q_{\text{mode}}$ & $\alpha$ & $q_{\text{mode}}$ & $\alpha$ & $q_{\text{mode}}$ & \\
    }
    \colnumbers
    \startdata
        IC~2391                 & 0.15 & 1.69 & $0.25^{+0.14}_{-0.12}$ & 0.42 & $0.33^{+0.14}_{-0.13}$ & 0.55 & $0.32^{+0.12}_{-0.10}$ & 0.40 & $0.20^{+0.17}_{-0.12}$ & 0.83 & 0.17$\div$0.20 \\
        IC~2602                 & 0.20 & 1.78 & $0.22^{+0.08}_{-0.06}$ & 0.38 & $0.20^{+0.11}_{-0.08}$ & 0.43 & $0.26^{+0.10}_{-0.08}$ & 0.33 & $0.14^{+0.08}_{-0.08}$ & 0.72 & 0.12$\div$0.21 \\
        Melotte~111 (Coma)      & 0.20 & 1.43 & $0.29^{+0.10}_{-0.12}$ & 0.73(?) & $0.36^{+0.17}_{-0.12}$ & 0.83(?) & $0.28^{+0.08}_{-0.14}$ & 0.55 & $0.26^{+0.15}_{-0.13}$ & 0.95 & 0.02$\div$0.05 \\
        Melotte~20 (Alpha Per)  & 0.22 & 2.43 & $0.21^{+0.06}_{-0.05}$ & 0.63 & $0.20^{+0.07}_{-0.05}$ & 0.57 & $0.21^{+0.07}_{-0.06}$ & 0.65(?) & $0.19^{+0.06}_{-0.08}$ & 0.70(?) & 0.05$\div$0.08 \\
        Melotte~22 (Pleiades)   & 0.20 & 1.30 & $0.29^{+0.05}_{-0.05}$ & 0.47 & $0.19^{+0.06}_{-0.06}$ & 0.67 & $0.29^{+0.05}_{-0.05}$ & 0.47 & $0.16^{+0.06}_{-0.05}$ & 0.85 & 0.03$\div$0.04 \\
        Melotte~25 (Hyades)     & 0.15 & 1.59 & $0.44^{+0.13}_{-0.10}$ & 0.60 & $0.33^{+0.10}_{-0.10}$ & 0.73 & $0.42^{+0.13}_{-0.12}$ & 0.65 & $0.32^{+0.10}_{-0.10}$ & 0.65 & 0.06$\div$0.12 \\
        NGC~2451A               & 0.17 & 1.68 & $0.28^{+0.08}_{-0.08}$ & 0.45 & $0.16^{+0.15}_{-0.11}$ & 0.62 & $0.30^{+0.11}_{-0.07}$ & 0.35 & $0.05^{+0.10}_{-0.04}$ & 0.83 & 0.16$\div$0.23 \\
        NGC~2632 (Praesepe)     & 0.22 & 1.69 & $0.29^{+0.11}_{-0.08}$ & 0.45 & $0.24^{+0.06}_{-0.06}$ & 0.83 & $0.29^{+0.08}_{-0.08}$ & 0.45 & $0.13^{+0.05}_{-0.04}$ & 0.83 & 0.01$\div$0.03 \\  
    \enddata
    \tablecomments{
        Full sample results are based on 50\%-membership photometric members. 
        Truncated sample excludes sources identified as likely false matches due to source confusion in cross-matching Gaia, 2MASS, and WISE catalogs. 
        The last column estimates the fractional loss of sources after truncation. Despite this loss, key trends in $\alpha$ and $q_{\text{mode}}$ remain robust.
        (?) means that the distribution has an uneven form for inference.}
\end{deluxetable*}

\section{The effects of different catalogs' resolution} \label{sec:Chance_matches}
False matches could affect count results, because they can contain unresolved binaries. 
To assess whether false matches impact the results significantly, we test samples without false matches that we identified during catalogs' cross-correlation.
The entire methodology was applied similarly in all mass intervals. 
However, this re-surveying required modifying HDBSCAN parameters for several clusters. We could keep the same initial parameters for Melotte~111, Melotte~20, Melotte~22, and NGC~2632. 
However, in the cases of IC~2391, IC~2602 and NGC~2451A, the relative content of sources in the purified samples is lower, as can be seen in column (12) of Tab.~\ref{tab:binary_results}. These values in column (12) should be interpreted as the fraction of systems by which the samples were reduced after excluding false matches.  
In addition, Melotte~25 has quite a spread main sequence.
Due to all these factors, we found that the modifications \texttt{min\_cluster\_size}$^*1.75$ and \texttt{min\_samples}$^*1.25$ are more appropriate.

The results presented in Tab.~\ref{tab:binary_results} for both cases of members samples are consistent within the limits of estimated errors. It shows that for the eight open clusters considered in this study, the sample discrepancies due to false matches are low. 
This could be explained by the fact that these open clusters are located close to the Sun and are therefore bright.
In case of further clusters, this effect might  play a more significant role.

\section{Method Reliability} \label{sec:Method_reliability}

To evaluate the applicability of our photometric method, we developed a series of numerical tests. 
The primary objective was to demonstrate that our data reduction system accurately retrieves the underlying mass ratio distribution $f(q)$ without introducing systematic artifacts, ensuring that a flat input distribution is recovered as flat and a Gaussian input is recovered as Gaussian. 

For this purpose, we generated synthetic star clusters for samples of 1000 sources (both singles and binaries), utilizing a three-segment Initial Mass Function (IMF) of \cite{Kroupa2001MNRAS} for primary masses and PARSEC isochrone with Pleiades parameters as a reference for magnitude-mass relation. 
Mass sampling is bounded within the $0.2-1.3 M_{\odot}$ interval, consistent with the processed membership of the Pleiades cluster. 
Our validation framework employed two distinct initial $f(q)$ distributions for secondary component masses with the true binary fraction fixed at 24\%: a uniform 'Flat` distribution across the  interval [0,1] and a 'Gauss` distribution peaked at $q=0.5$ --- a characteristic of observed clusters. 
To distinguish the intrinsic properties of the cluster and the expected impact of observational noise, we generated two sets of models for each scenario: one error-free and one that account for magnitude-dependent photometric noise. 
The recovery of $f(q)$ was then performed using our pipeline, subtracting histograms and box-and-whisker plot.

A critical feature of our methodology, as presented in the final visualizations, is the use of a hybrid box-and-whisker plot construction designed to prioritize the observed properties of the clusters. 
In these plots, the central markers (orange and white dashes) represent the results of the error-free model samples. 
To this deterministic baseline, we 'attach` IQR and whiskers derived from 100 Monte Carlo realizations where photometric errors were varied in error-polluted samples. 

\begin{figure}[h!]
    \centering
    \includegraphics[width=0.9\linewidth]{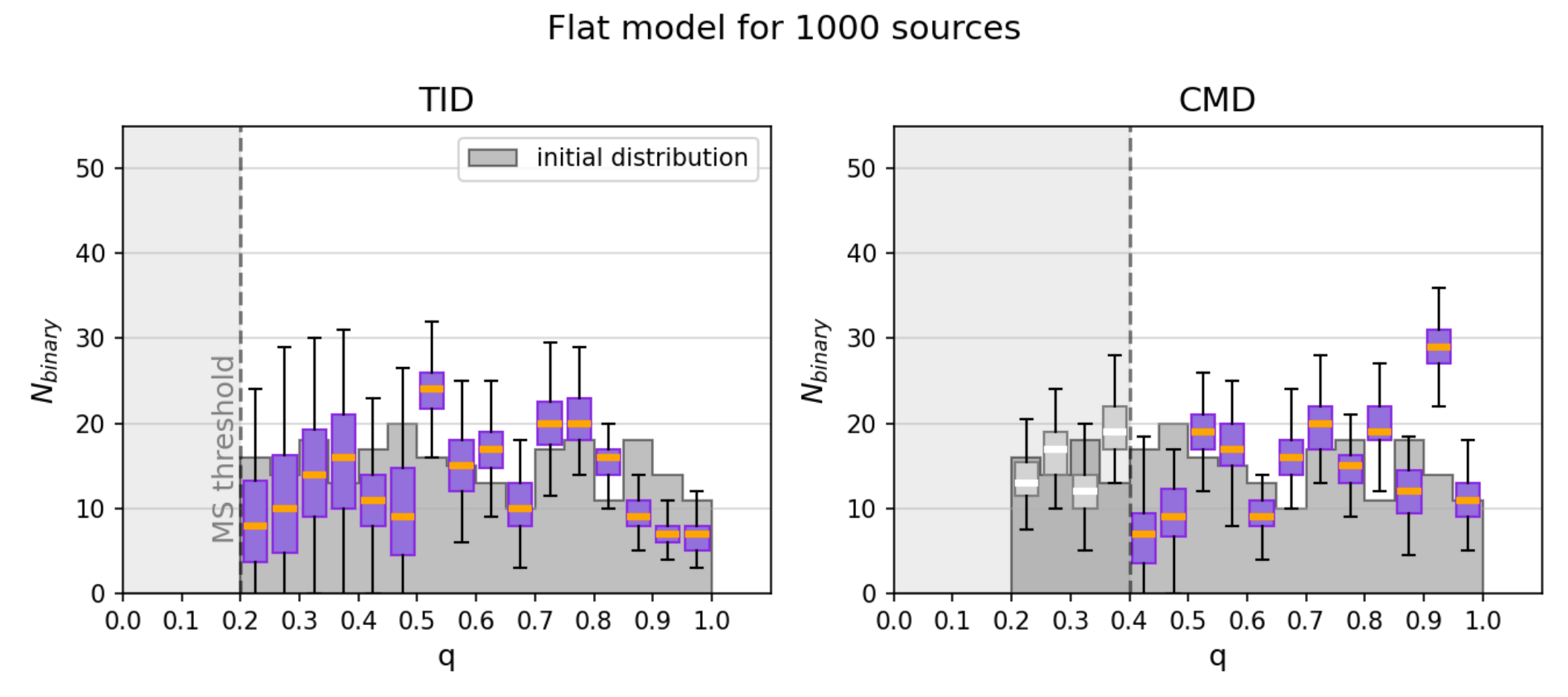}
    \includegraphics[width=0.9\linewidth]{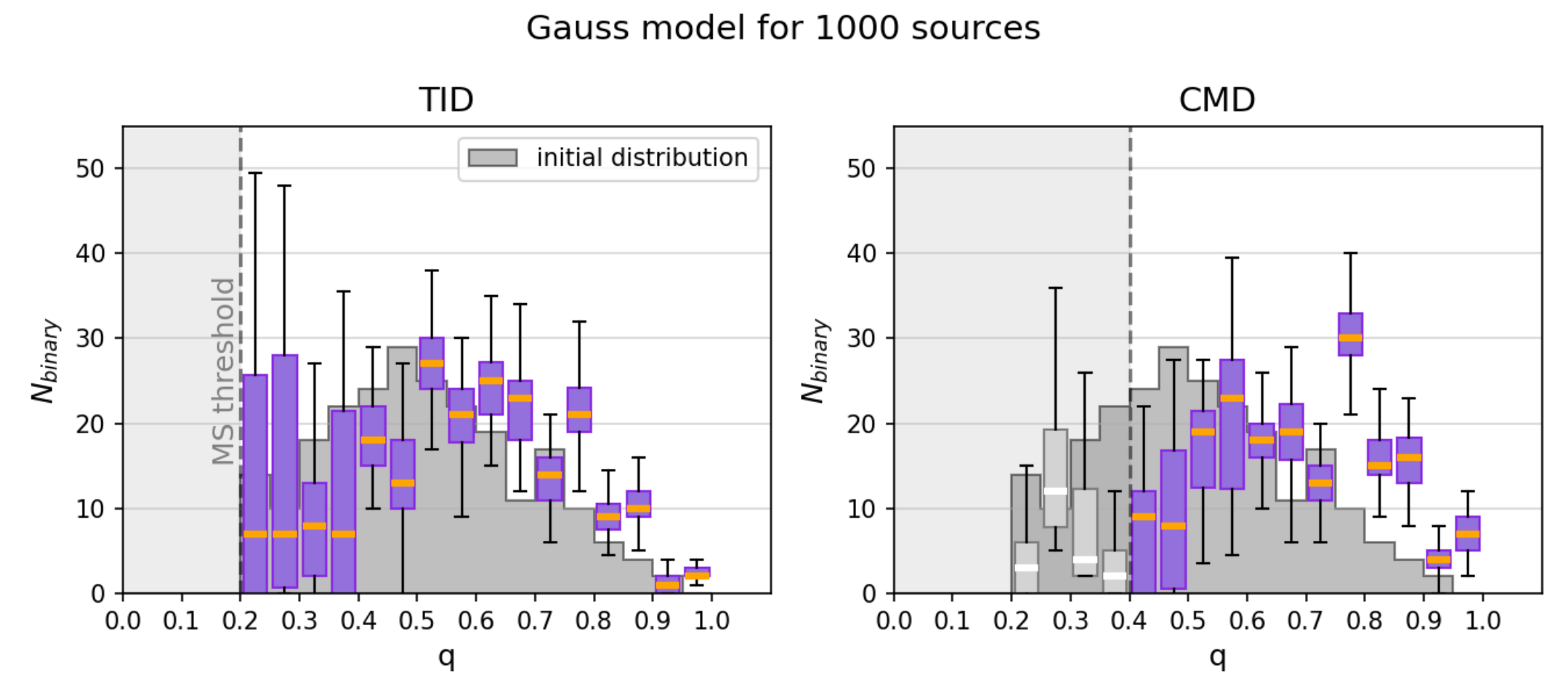}
    \caption{Recovered distribution for the Flat and Gauss models of 1000 sources. Orange dashes present values from sample unpolluted by photometric errors. The purple boxes depict IQRs obtained from variations of this model sample with photometric errors.}
    \label{fig:model_recovery}
\end{figure}

To quantify the global agreement between the recovered and input mass-ratio distributions shape, chi-square for normalized histograms ($\chi_{shape}^2$) and Wasserstein distance ($W_1$) metrics were used.
$\chi_{shape}^2$ is calculated using histograms normalized to unity. It is sensitive to discrepancies in individual bins.
$W_1$ is a metric to measure the cumulative mass (probability) difference across the q grid. 
In our case it quantifies how much the recovered cumulative binary fraction deviates from the model, integrated over q range.

The results of simulations processing confirm the robustness of our analysis system and our considerations about usage purely visual data for unresolved binaries search. 
Binary fraction could be recovered within the estimated errors range. 
The statistical criteria indicate that the recovered $N_{binary}(q)$ is statistically consistent with the input model shapes $f(q)$ in both TID and CMD spaces but it is better for TID. 
This is presented in Table~\ref{tab:stat_criteria_for_models}. 

Our method demonstrates fidelity in retrieving input distributions, but we recognize that systematic uncertainties may arise from theoretical isochrone shifts. 
The probability of a system being recovered in a specific bin depends on the local isochrone spacing and q-bin width. 
Discrepancies between PARSEC models and observed sequences can result in misplaced isolines. 
The non-uniform spacing of theoretical q-intervals acts as a statistically preferred binning geometry, potentially inducing apparent peaks in the recovered f(q).
In TID space, the narrow bins at $q\approx0.9\div1.0$ contrast with the broader intervals at $q\approx0.6\div0.8$. Photometric noise scatters systems into these intervals, creating an artificial peak near $q\approx0.7$. Conversely, the CMD high-$q$ bins ($q\approx0.7\div0.9$) are wider, leading to a systematic shift towards $q\approx1$. 

These features are binning and error-influenced artifacts. It must be noticed that the recovered shape of any mass-ratio distribution is fundamentally dependent on the photometric combination employed. However, since the observed peak ($\sim0.4-0.6$ for TID and $\sim0.5
-0.8$ for CMD) is distinct from the primary artifact peak, this suggests the observed peak is physical.

\begin{deluxetable}{lcccc}
    \tabletypesize{\scriptsize}
    \label{tab:stat_criteria_for_models}
    \tablewidth{\linewidth} 
    \tablecaption{Quantitative Shape Metrics: Chi-Square and Wasserstein Distance}
    \tablehead{
    \multirow{2}{0.1\textwidth}{\shortstack{Modeled \\ $f(q)$}} & Space & Criteria & Criteria Values & \multirow{2}{*}{\shortstack{Binary Fraction \\ $\alpha$}} \\
     }
    \colnumbers
    \startdata 
        \multirow{4}{0.1\textwidth}{Flat} & \multirow{2}{*}{TID} & Shape chi-square & $\chi_{shape}^2$=0.16, $\nu$=15, $p$=1 &  \multirow{2}{*}{$0.24^{0.06}_{0.05}$} \\ 
         & & Wasserstein distance & $W_1$=0.0273 & \\ 
         & \multirow{2}{*}{CMD} & Shape chi-square & $\chi_{shape}^2$=0.20, $\nu$=15, $p$=1 &  \multirow{2}{*}{$0.25^{0.04}_{0.03}$} \\ 
         & & Wasserstein distance & $W_1$=0.0428 & \\ \cline{1-5}
        \multirow{4}{0.1\textwidth}{Gauss} & \multirow{2}{*}{TID} & Shape chi-square & $\chi_{shape}^2$=0.35, $\nu$=14, $p$=1 & \multirow{2}{*}{$0.23^{0.10}_{0.06}$} \\ 
         & & Wasserstein distance & $W_1$=0.0763 & \\ 
         & \multirow{2}{*}{CMD} & Shape chi-square & $\chi_{shape}^2$=1.20, $\nu$=14, $p$=1 & \multirow{2}{*}{$0.20^{0.06}_{0.06}$} \\ 
         & & Wasserstein distance & $W_1$=0.1039 &  \\
    \enddata
    \tablecomments{The p-value is the probability of observing a test statistic at least as extreme in a chi-squared distribution assuming that the null hypothesis is true. The null hypothesis states that observed frequencies match expected ones; low $\chi_{shape}^2\ll\nu$ with $p\approx1$ confirms shape match. A value of $W_1=0$ implies identical shapes, while larger values indicate a stronger shift of weight along the q-scale; $W_1\ll0.1$ suggest good agreement as the q-scale ranges from 0 to 1.}
\end{deluxetable}

\vfill
\section{Discussion} \label{sec:Discussion}
This paper presents a detailed study on the identification and analysis of unresolved binary star systems in eight open star clusters located within 200 parsecs of the Sun. 
We introduce empirical isochrones that enhance the accuracy of estimating the binary star fraction and allow us to derive the distribution of the component mass ratio $q$.

We exploited the (H-W2)-W1 vs W2-(BP-K) photometric diagram along with Gaia photometry to identify and analyze binary systems. 
We demonstrated that TID is more effective in separating single and binary stars, even for systems with small mass ratios $q<0.5$, compared to the routinely employed CMD.
This is due to usage of the infrared bands because the low-mass component exhibits some infrared excess and makes unresolved binaries redder.
In addition, Figure~\ref{appendix_fig:ms_err} in Appendix~\ref{appendix:ms_errors} demonstrates how photometric errors and the spread of MS single stars in the CMDs produce contamination in low-q intervals, thus limiting the use of the CMD for binaries detection. These errors also set limits on the implementation of the methodology, as illustrated by dashed lines in Fig.~\ref{appendix_fig:qDistr_Whisker}.

\begin{figure}[h!]
    \centering
    \includegraphics[width=0.9\linewidth]{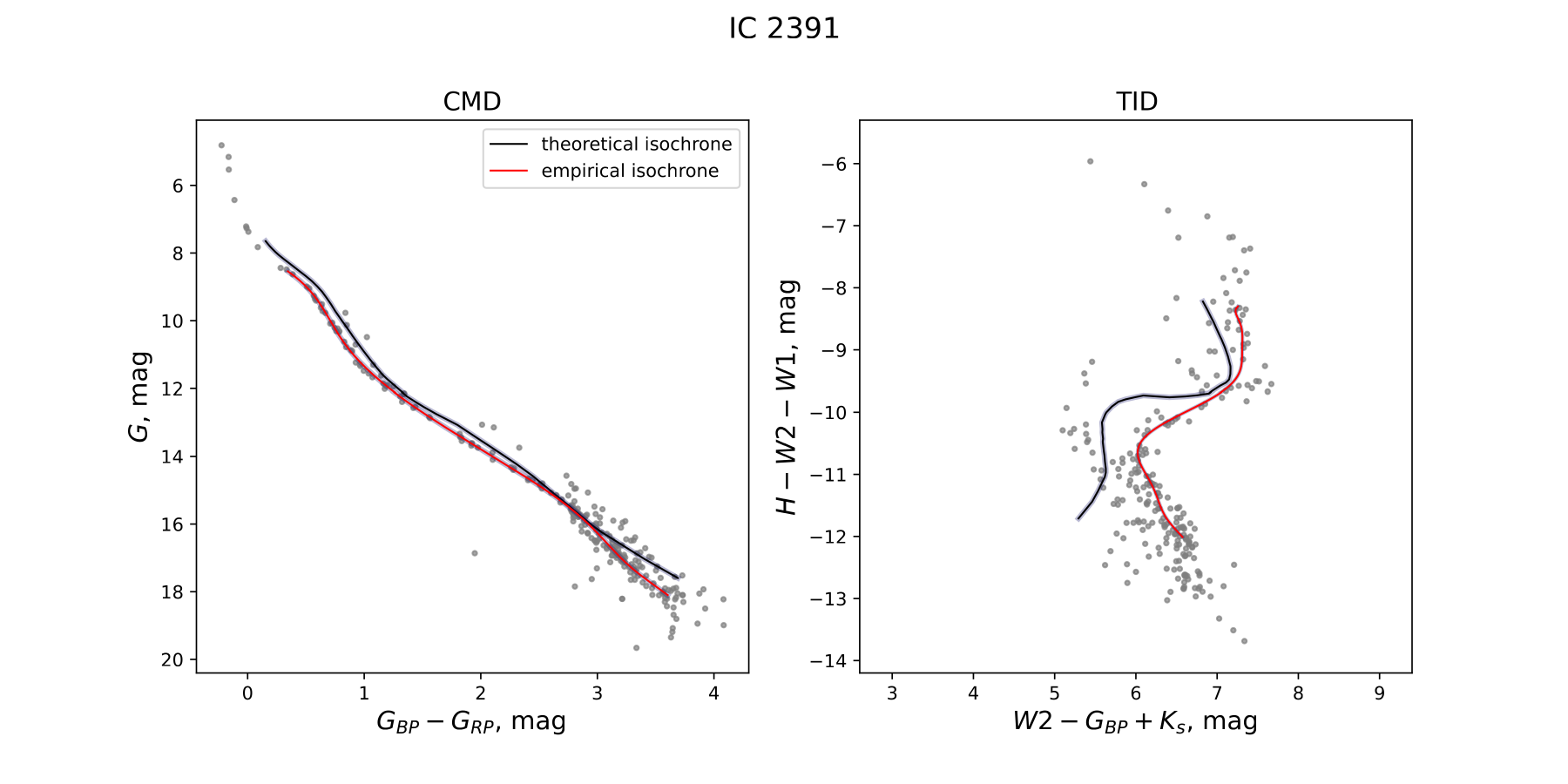}
    \caption{CMD and TID of IC~2391 with an empirical isochrone (red line) and a theoretical PARSEC isochrone (black line) with fundamental parameters from \citet{Dias+2021}.}
    \label{fig:empir_and_teor_iso}
\end{figure}

The HDBSCAN algorithm is used to identify single stars in CMD and TID, which helps in constructing empirical isochrones for single and binary stars. The introduction of the empirical isochrones approach addresses the limitations of theoretical isochrones providing a more accurate representation of the main sequence (Fig.~\ref{fig:empir_and_teor_iso}). 
This approach allows for a wider operational primary mass interval, particularly in the low-mass region, where the secondary companion is considered to be mostly a brown dwarf less massive than $0.075 M_{\odot}$, as was assumed in \citet{Malofeeva+2023}. 

\subsection{On the Binary Fraction}
Our estimates of the binary fraction $\alpha$ lie within the range of 0.16-0.36 when using CMDs, and 0.21-0.44 when using TIDs. 
This suggests that in our previous studies we may have over-estimated the binary fraction as we improved the accuracy of the method due to introduction of empirical isochrones.

\begin{figure}[h!]
    \centering
    \includegraphics[width=0.6\linewidth]{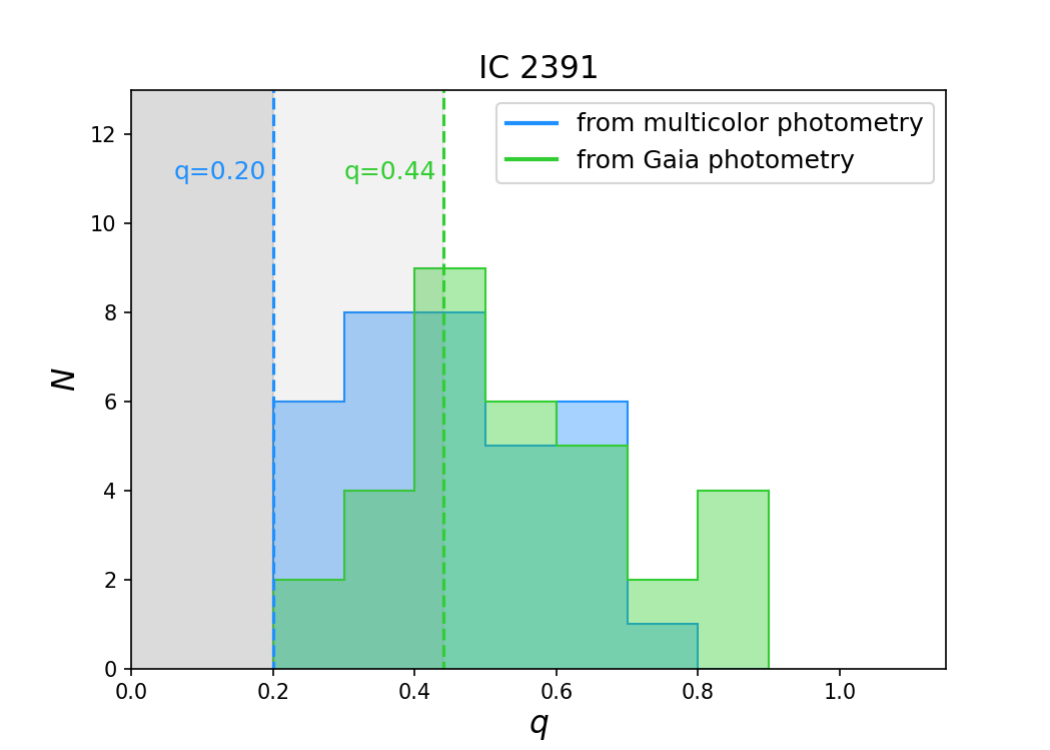}
    \caption{Distributions of binary stars with confidence intervals obtained from multicolor data (blue line) and Gaia data (green line). Dashed lines denote the limits of results reliability according to photometrical errors (Fig.~\ref{appendix_fig:ms_err}). Shadowed with dark gray (multicolor) and light gay (Gaia) regions should not be considered in the analysis.}
    \label{fig:comparative_diagram}
\end{figure}

Many recent studies have been devoted to the properties of binary stars in OCs, including the binary fraction and the distribution of binary mass ratio. 
However, their methods could be applied only for $q>0.5$. Binary fractions range from 0.06 to 0.8, with a median of 0.17--0.18 \citep{Donada2023A&A,Jiang2024ApJ}. 
\cite{Niu2020ApJ} report about 0.29 -- 0.55, while \cite{Liu2023ApJS} write about 0.09 -- 0.44. As seen, our current estimations are consistent with other works, but more complete in the sense of the mass-ratio diversity (our limit is $q>0.2$). It could also be recognized from the binaries distribution in Fig.~\ref{fig:comparative_diagram} 
, where the capability of TID in low-q binaries detection is vividly illustrated.

\subsection{On the Mass Ratio Distribution}
The mass ratio distribution of binary stars is an important parameter in understanding stellar multiplicity and cluster dynamics.
In our study the mode of the distributions is found to be in the range of 0.43-0.83 for Gaia photometry. 
For infrared-visible photometry it is 0.38-0.63, widening the interval of generally proposed values and implying a notable low-mass-ratio binaries contribution in the fraction. 

\begin{figure}[h!]
    \centering
    \includegraphics[width=0.5\linewidth]{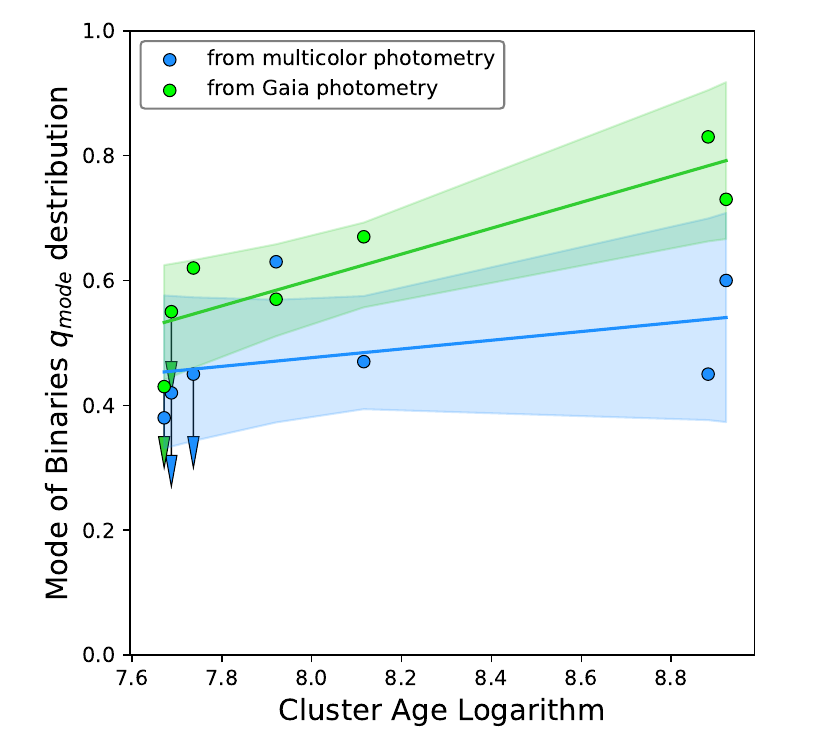}
    \caption{The relation between $q_{mode}$ of binaries distribution and clusters' age with linear regression (Eq.~\ref{eq:regressions}) and confidence intervals. Several $q_{mode}$-values might be less then it shown on the figure due to methodology limitation. It is denoted by arrows.}
    \label{fig:age_vs_qmode}
\end{figure}

That addresses the challenge of low-q binaries detection.
We found that $q_{mode}$ increases in older clusters as shown in Fig.~\ref{fig:age_vs_qmode} with the following law:

\begin{equation}
    \begin{array}{c}
        y= x \cdot (0.07\pm 0.07) - (0.08\pm 0.57)\,\text{ for TID},\\
        y= x \cdot (0.21\pm 0.05) - (1.06\pm 0.43)\,\text{ for CMD}.
    \end{array}
    \label{eq:regressions}
\end{equation}

Remarkably, in IC~2391, IC~2602 clusters $q_{mode}$ might turn out to be less then it was extracted from Gaia data. That also applies for IC~2391 and NGC~2451A from multicolor data.
That is because the boundaries of the methodology implementations, denoted with dashed lines in Fig.~\ref{appendix_fig:qDistr_Whisker}, close to maximums of distributions. 
As the distributions from the multicolor photometry increase toward the lower $q$, we assume that $q_{mode}$ could be lower and denote it with arrows in Fig.~\ref{fig:age_vs_qmode}. Thus, the regression estimations can be steeper.

Different authors discuss the mass-ratio distribution in a number of contexts. \cite{Cordoni2023A&A} and \cite{Niu2020ApJ} found that the mass-ratio distribution of binaries is generally flat, with no significant variation with the primary mass or mass ratio. 
\cite{Kouwenhoven+2009} suggested the Gauss law for describing the q distribution and in \cite{Malofeeva+2023} it was implemented in agreement with the data. 
Others find evidence of a power-law distribution with significant uncertainties \citep{Chulkov2025AJ}. 

We found consistency with previous studies, since a power law is not suitable because the distributions have a maximum. 

However, it is worth noting that a decrease in the left side of both `blue' and `green' distributions in Fig.~\ref{fig:comparative_diagram} are not reliable due to method limitations ($q_{TID}>0.2$, $q_{CMD}>0.5$). This region is shadowed with gray color. As a consequence, one could end up suggesting a larger fraction of low-q systems.
We also recall that, according to \cite{Liu2025AJ}, the mass-ratio distribution follows a three-segment power-law profile with a peak at $q\gtrsim0.95$.

\subsection{On the Twin-Excess}

The excess of binaries with equal mass components has been reported in several more studies: \cite{El-Badry+2019,Alexander2025MNRAS}.
In this study, we confirm our previous findings that twin-binaries are not a preferred population.

A comparison with recent work by \cite{Childs2025arXiv}, who analyzed 35 open clusters using a Bayesian isochrone-fitting method (BASE-9), reveals an apparent discrepancy: they report mass-ratio distributions that are largely flat or tend toward higher mass-ratio values with no universal preference for twin systems, but report about a peak near $q=1$ for old, dynamically evolved clusters. However, their analysis is limited to binaries with $q>0.5$ , as lower-q systems are difficult to distinguish from single stars in their modeling framework (see \cite{Childs2024ApJ,Cohen2020AJ}).

In contrast, our multiband Two-Index Diagram (TID) approach enhances sensitivity to low-q binaries by leveraging color shifts in infrared-optical composite indices, allowing reliable detection down to $q\sim0.2$. The rising trend in $f(q)$ towards lower mass ratios, seen clearly in Figures \ref{fig:comparative_diagram}, suggests a rich population of unequal-mass binaries that fall below the completeness threshold of methods like BASE-9.

Thus, the two studies likely probe different parts of the mass-ratio spectrum: ours emphasizes the low-q end, revealing a peak in the distribution, while \citet{Childs2025arXiv} focus on the high-q regime, where the distribution appears flatter or mildly enhanced at unity. This complementarity highlights the importance of multi-method approaches in reconstructing the full binary mass-ratio function.

Moreover, dynamical evolution may play a role. \citet{Childs2025arXiv} find that the median $q$ increases with the dynamical age of the cluster, suggesting that interactions increase the proportion of binaries of high-$q$ over time. Our sample consists predominantly of nearby, relatively young clusters (ages $\lesssim 1$ Gyr), which may better preserve primordial $f(q)$ shapes skewed toward low $q$, consistent with star formation models predicting preferential formation of unequal-mass pairs (e.g., \citet{Kroupa1993MNRAS,Bate2012MNRAS}).

These findings are also consistent with high-precision spectroscopic studies. \citet{Geller2021AJ} analyzed the old open cluster M67 (age $\approx 4$ Gyr) and found that its completeness-corrected mass-ratio distribution is consistent with a flat $f(q)$ across $0.1<q<1$, with no strong preference for twin systems. In contrast, for the older, more dynamically relaxed NGC 188 (age $\approx 6$ Gyr), they observed an increase in $f(q)$ at high mass ratios, culminating in a local peak near $q=1$. The assumption behind this attribute is that twin-rich populations are not primordial, but rather the result of long-term dynamical processing where low-q and wide binaries are more frequently disrupted or evaporated from the core.

These results suggest that, while primordial star formation may favor unequal-mass pairs (as seen in our young clusters), subsequent dynamical interactions can reshape the $f(q)$ distribution, increasing the proportion of high-q binaries over time. This evolutionary picture is reinforced by \citet{Childs2025arXiv}, who found a positive correlation between the median $q$ and the dynamical age of the group in 35 systems. The flat distribution seen in M67 by \citet{Geller2021AJ} does not contradict our findings, because M67 ($\sim4$ Gyr) lies at the extreme end of dynamical evolution compared to our sample, which is dominated by clusters younger than 1 Gyr. The peaked distributions, centered on the $q\sim0.4-0.6$ distributions, we observe are likely representative of the initial conditions established during star formation, before significant encounters reshape the binary population.

Together, these results suggest a coherent evolutionary sequence: young clusters form with a preponderance of unequal-mass binaries; over time, dynamical interactions deplete the low-q and wide population, leading to flatter or twin-enhanced distributions in older systems like M67 and NGC 188. Our work focuses on capturing the early phase of this process, leveraging public photometry to map close to primordial multiplicity trends across many clusters uniformly.

The completeness correction applied by \citet{Geller2021AJ} accounts for observational biases in radial velocity surveys (e.g., period coverage, signal-to-noise limitations) using Monte Carlo simulations calibrated on known binaries. Although robust, such corrections depend on assumed prior distributions and remain limited to clusters with long-term spectroscopic monitoring, rare exceptions rather than the norm.

Our photometric approach offers a complementary perspective: less precise per system, but uniformly applicable across hundreds of clusters using only public imaging data. By combining optical and infrared photometry in TID, we achieve sensitivity down to $q\sim0.2$, enabling population-level statistics even in the absence of spectroscopy.

Our finding that $f(q)$ peaks at moderate mass ratios and declines toward q=1 appears to be at odds with the classical notion of a distinct `binary sequence' located approximately 0.75 mag above the single-star main sequence in color-magnitude diagrams (CMDs). However, this feature does not necessarily imply a dominant population of twin binaries ($q\approx1$). Modern analyses show that the upper MS is often broad and continuous, consistent with a superposition of unresolved binaries across a wide range of $q$, rather than a narrow sequence of twins \citep{Niu2020ApJ}. Moreover, other astrophysical phenomena --- such as blue stragglers, rapid rotators, and hierarchical triples --- also populate this region \citep{Muratore2024A&A}, further diluting any pure twin signal.

The apparent disagreement with \cite{Raghavan+2010}, who report a flat $f(q)$ with a modest excess at $q=1$ for solar-type field stars within 25 pc, can be resolved through dynamical evolution. Field binaries originate from a variety of progenitor environments, including old open clusters such as M67 and NGC 188, where dynamical interactions favor the survival and hardening of high-q systems \citet{Geller2021AJ}. It is the contribution of old clusters that predominates in the field star population. The twin excess observed by \citet{Raghavan+2010} may therefore reflect the cumulative effect of cluster dissolution over billions of years, rather than the initial conditions of binary formation. 

In contrast, our sample focuses on young ($<1$ Gyr), nearby clusters that are less affected by internal dynamics. The peaked $f(q)$ distribution that we observe --- centered at $q\sim0.4-0.6$ --- likely represents the primordial mass-ratio function, consistent with hydrodynamic simulations of star formation that favor fragmentation into unequal-mass cores \citet{Bate2012MNRAS}.

Finally, we note that detecting twins photometrically is inherently challenging: in Gaia CMDs, the photometric shift of a $q\sim1$ binary is nearly parallel to the main sequence, making it difficult to separate from binaries with $q$ down to 0.8. While our multiband Two-Index Diagram (TID) improves distinction, there remains an intrinsic degeneracy in broadband photometry: systems with very similar mass ratios (e.g., $q>0.8$ ) or different astrophysical origins (e.g., blue stragglers, rapid rotators) can occupy overlapping regions in the diagram. This is a fundamental physical and observational limit, even if the overall detection efficiency is improved.

\subsection{On the Pleiades}
We now focus on the Pleiades cluster, which has been the subject of intense investigation. Previous studies report binaries fraction ranging from 0.06 to 0.41 \citep{Cordoni2023A&A,Donada2023A&A,Niu2020ApJ,Liu2025AJ,Jadhav2021AJ,Chulkov2024AJ,Liu2023ApJS}. 
We find $\alpha=0.29\pm0.05$ in TID and $\alpha=0.19\pm0.06$ in CMD. 
We favor the two-index diagram method because it provides more robust detection of unresolved binaries and its fraction estimate. 
Our research finds no significant impact of false close neighbors (due to catalog cross-matching) on the accuracy of binary fraction estimates. 
We ensure minimal sample discrepancies, thus providing a robust dataset for analysis.

Note that samples of probable cluster members usually contain sources with well-determined astrometric solutions. Therefore, stars with bad RUWE ($>1.4$) or with a two-parametric solution are absent in the probable cluster member samples. According to \cite{Tagaev2025}, about half of the probable members could be lost in the case of NGC~3532. Many unresolved binaries could instead be present among these lost stars.

We considered the known spectroscopic binaries in the Pleiades region according to the catalog of \cite{Torres2021ApJ}. 
We found 18 double-lined spectroscopic binaries (SB2) and, among them, 13 ones could be analyzed using our methodology. All these sources are located in the unresolved binaries' region in diagrams. 
We compared our $q$ estimations with the spectroscopic binaries data sets in Tab.~\ref{tab:comparison_with_SB2}. 
As seen, our results show reasonable agreement with the spectroscopic data, excluding sources HII~164, HII~2406, and PELS~38.

The secondary components in HII~164 and PELS~38 are very faint according to \cite{Torres2021ApJ}, and its orbit solution is considered preliminary, so the orbital elements (such as the semi-amplitudes of the velocity $K_1$, and $K_2$, from which $q=K_1/K_2$) are not trusted. HII~2406 is a young stellar object by \citet{Kounkel2019AJ}, and the note in the catalog of \cite{Torres2021ApJ} suggests that its solution is as reliable as HII 164. Hence, we do not have solid information for these sources to compare.
We also suggest two sources (HII~320 and HCG~489) as multiples.
\cite{Torres2021ApJ} report that the secondary component HII~320 is a rapidly rotating star. Consequently, it increases reddening of a pair and we assign this source as multiple because of it's location in TID. 

We consider HCG~489 as multiple system, but it is very close to the boundary between binaries and multiples.
Then, it is possible to classify that as a twin-binary star as well. As a result, multicolor photometry allows us to estimate the mass ratio in an unresolved binary system as reliably as spectroscopy.

\begin{deluxetable}{lccc}
    \label{tab:comparison_with_SB2}
    \tablewidth{0pt} 
    \tablecaption{Comparison of our $q$ estimations with spectroscopic binaries data from \cite{Torres2021ApJ} in Pleiades.}
    \tablehead{
    Source & $q_{SB2}$ & $q_{TID}$ & $q_{CMD}$\\
    }
    \colnumbers
    \startdata 
        AK I-2-288 & 0.87$\pm$0.0053 & 0.74$\pm$0.0081 & 0.84$\pm$0.0427 \\
        DH 794 & 0.71$\pm$0.0085 & 0.74$\pm$0.0263 & 0.75$\pm$0.0001 \\
        HII 164 & 0.32$\pm$0.017 & $< 0.2$ & $< 0.44$ \\
        HII 173 & 0.95$\pm$0.011 & 0.75$\pm$0.0002 & 0.84$\pm$0.0465 \\
        HII 320 & 0.86$\pm$0.023 & multiple & 0.84$\pm$0.0292 \\
        HII 761 & 0.68$\pm$0.007 & 0.65$\pm$0.0392 & 0.55$\pm$0.0434 \\
        HII 1117 & 0.98$\pm$0.014 & 0.65$\pm$0.014 & 0.75$\pm$0.0001 \\
        HII 1348 & 0.78$\pm$0.0098 & 0.76$\pm$0.0568 & 0.75$\pm$0.04 \\
        HII 2406 & 0.54$\pm$0.017 & 0.25$\pm$0.0027 & $< 0.44$ \\
        HCG 384 & 0.82$\pm$0.035 & 0.95$\pm$0.0001 & 0.96$\pm$0.0392 \\
        HCG 489 & 0.99$\pm$0.0034 & multiple & multiple \\
        HCG 495 & 0.98$\pm$0.012 & 0.75$\pm$0.0376 & 0.96$\pm$0.0559\\
        PELS 38 & 0.57$\pm$0.012 & 0.45$\pm$0.0497 & 0.45$\pm$0.0172
    \enddata
\end{deluxetable}

\subsection{Methodology Limitations}
The most important limitation of our investigation remains isochrones. The suggested approach improves reliability, but keeps available only low and intermediate masses for the primary component (see Tab.~\ref{tab:binary_results}, columns (2)-(3)). The clustering algorithm is unable to distinguish the condensation of the main sequence if the star cluster contains relatively few members or the cluster sequence looks loose. This also applies to the top of the cluster sequence, where it becomes sparse. 
But this part of a sequence is also brighter. The multiplicity could be resolved by a speckle interferometry survey up to the magnitude limit $G<15$ \citep{Chulkov2025AJ}. 
The binning geometry of the chosen photometric space also introduces systematic effects that must be carefully distinguished from intrinsic stellar properties.

\section{Conclusions} \label{sec:Conclusion}
We analyzed 8 open clusters using two independent approaches: CMD and TID. Our comparative study reveals that results derived from Gaia-only CMDs systematically underestimate the binary fraction and skew the mass ratio distribution toward higher $q$, due to limited sensitivity below $q\lesssim0.4$. In contrast, the multiband TID approach detects a significantly richer population of binaries with $q\sim0.2$, yielding a more complete census. While CMD-based studies remain valuable, our work demonstrates that combining optical and infrared photometry is essential for accurate characterization of unresolved binary populations in open clusters. To reach a lower limit in $q$, we introduced empirical isochrones that improve the fit of the cluster's main sequence compared to the theoretical isochrones. We were able to estimate the $q$-value for every star.
In addition, we found that the mode of the $q$-distribution increases with cluster age.

\begin{acknowledgments}
The work of V.O.Mikhnevich and A.F.Seleznev was supported by the Ministry of Science and Higher Education of the Russian Federation, FEUZ-2026-0012. 

This work were carried out with data from the European Space Agency (ESA) mission {\it Gaia} (\url{https://www.cosmos.esa.int/gaia}), processed by the {\it Gaia} Data Processing and Analysis Consortium (DPAC, \url{https://www.cosmos.esa.int/web/gaia/dpac/consortium}). Funding for the DPAC has been provided by national institutions, in particular the institutions participating in the {\it Gaia} Multilateral Agreement. We also make use of data products from the Two Micron All Sky Survey, which is a joint project of the University of Massachusetts and the Infrared Processing and Analysis Center/California Institute of Technology, funded by the National Aeronautics and Space Administration and the National Science Foundation. And the Wide-field Infrared Survey Explorer, which is a joint project of the University of California, Los Angeles, and the Jet Propulsion Laboratory/California Institute of Technology, and NEOWISE, which is a project of the Jet Propulsion Laboratory/California Institute of Technology. WISE and NEOWISE are funded by the National Aeronautics and Space Administration.
\end{acknowledgments}

\vspace{5mm}

\software{astropy \citep{astropy:2013, astropy:2018, astropy:2022}, HDBSCAN \citep{McInnes2017}, SAO/NASA Astrophysics Data System}

\newpage
\appendix
\section{HDBSCAN Parameters}\label{appendix:hdbscanPar}
\begin{deluxetable*}{lcccccc}[h!]
    \label{tab:hdbscan_par}
    \tablewidth{0pt} 
    \tablecaption{HDBSCAN parameters}
    \tablehead{
    \multirow{3}{*}{Cluster} & \multicolumn{4}{c}{TID} & \multicolumn{2}{c}{CMD} \\
    \cline{2-7}
     & \multicolumn{2}{c}{top sequence} & \multicolumn{2}{c}{bottom sequence} & \multirow{2}{*}{\texttt{min\_cluster\_size}} & \multirow{2}{*}{\texttt{min\_samples}}\\
    \cline{2-5}
     & \multicolumn{1}{c}{\texttt{min\_cluster\_size}} & \multicolumn{1}{c}{\texttt{min\_samples}} & \multicolumn{1}{c}{\texttt{min\_cluster\_size}} & \multicolumn{1}{c}{\texttt{min\_samples}} &  & 
    }
    \startdata 
        IC~2391 & 12 & 4 & 16 & 7 & 6 & 2 \\
        IC~2602 & 6 & 3 & 12 & 6 & 6 & 3 \\
        Melotte~111 & 6 & 3 & 12 & 6 & 3 & 1 \\
        Melotte~20 & 12 & 6 & 12 & 15 & 12 & 6 \\
        Melotte~22 & 14 & 5 & 28 & 4 & 20 & 10 \\
        Melotte~25 & 4 & 6 & 4 & 6 & 6 & 3 \\
        NGC~2451A & 8 & 3 & 15 & 3 & 6 & 3 \\
        NGC~2632 & 7 & 4 & 8 & 6 & 14 & 7 \\
    \enddata
\end{deluxetable*}

\newpage
\section{Empirical Isochrones}\label{appendix:EIs}
\begin{figure}[ht!]
    \centering
    \includegraphics[width=0.45\textwidth]{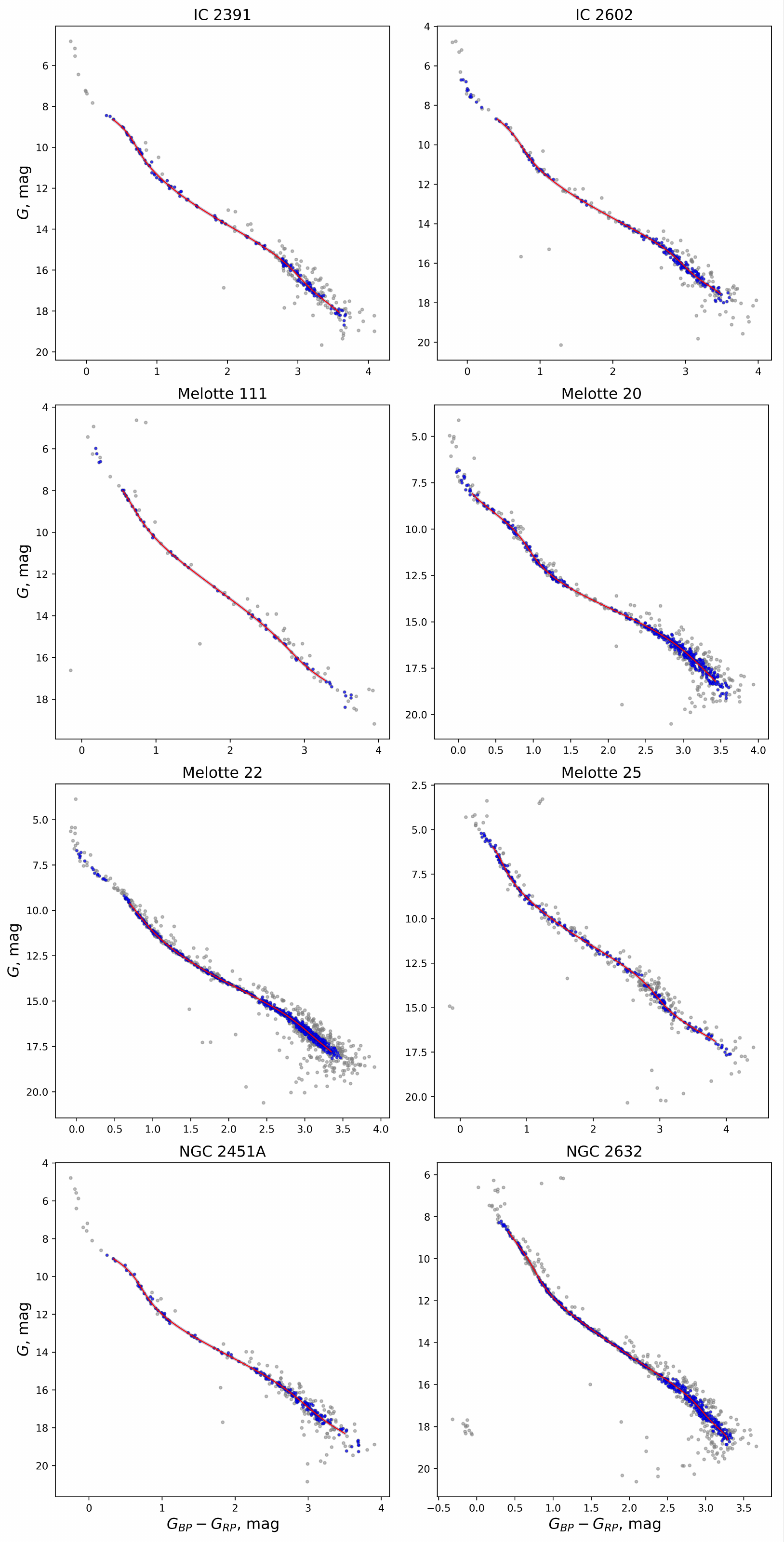}
    \includegraphics[width=0.45\textwidth]{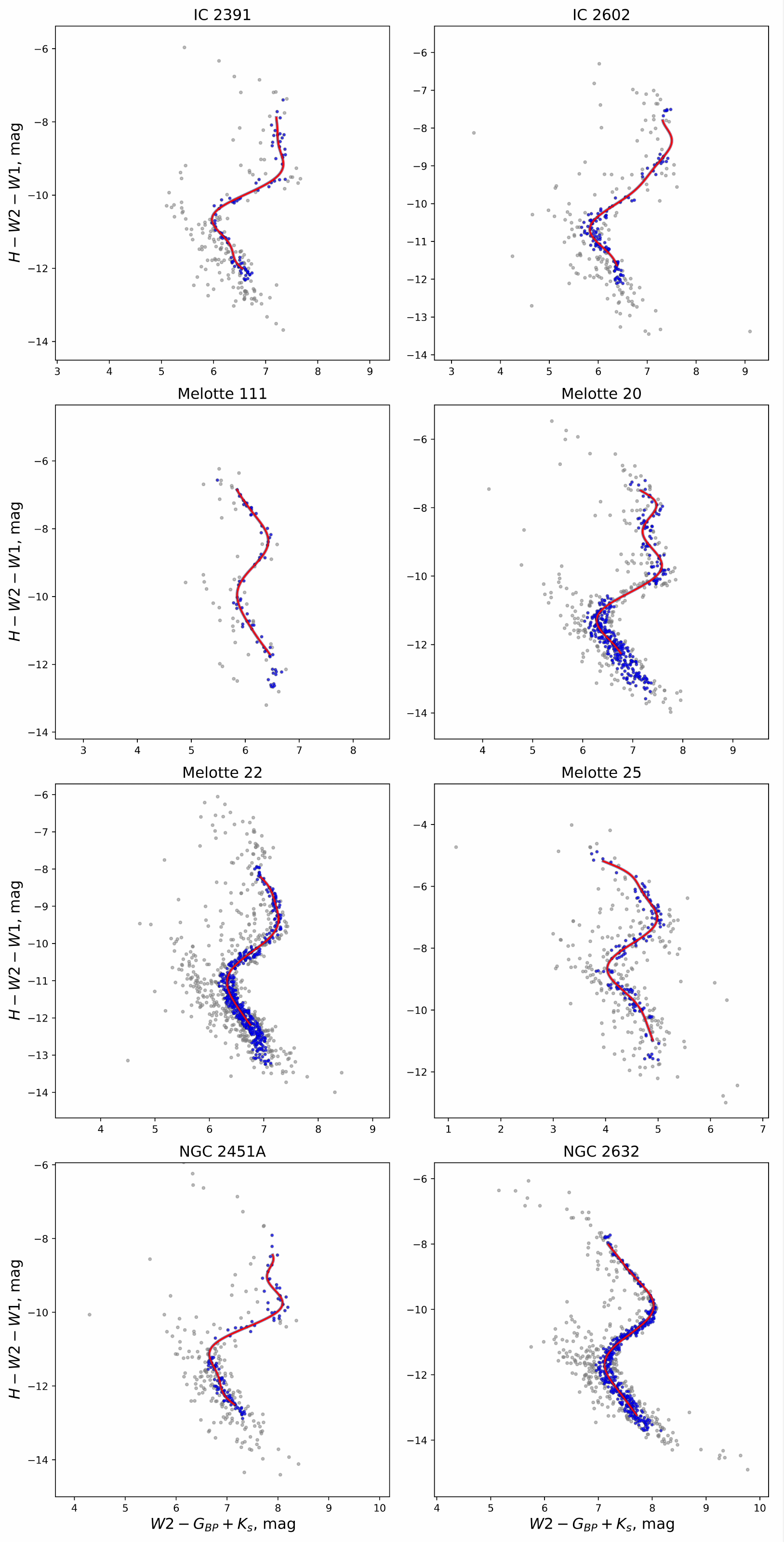}
    \caption{Empirical isochrones (red line) of single stars assigned by HDBSCAN (blue dots) presented on CMDs and TMDs. Grey points are probable cluster members from \cite{Hunt&Reffert2024} with $p>50\%$.}
    \label{appendix_fig:ei_1}
\end{figure}

\newpage
\section{Main Sequence Sources Errors}\label{appendix:ms_errors}
\begin{figure}[ht!]
    \centering
    \includegraphics[width=0.45\textwidth]{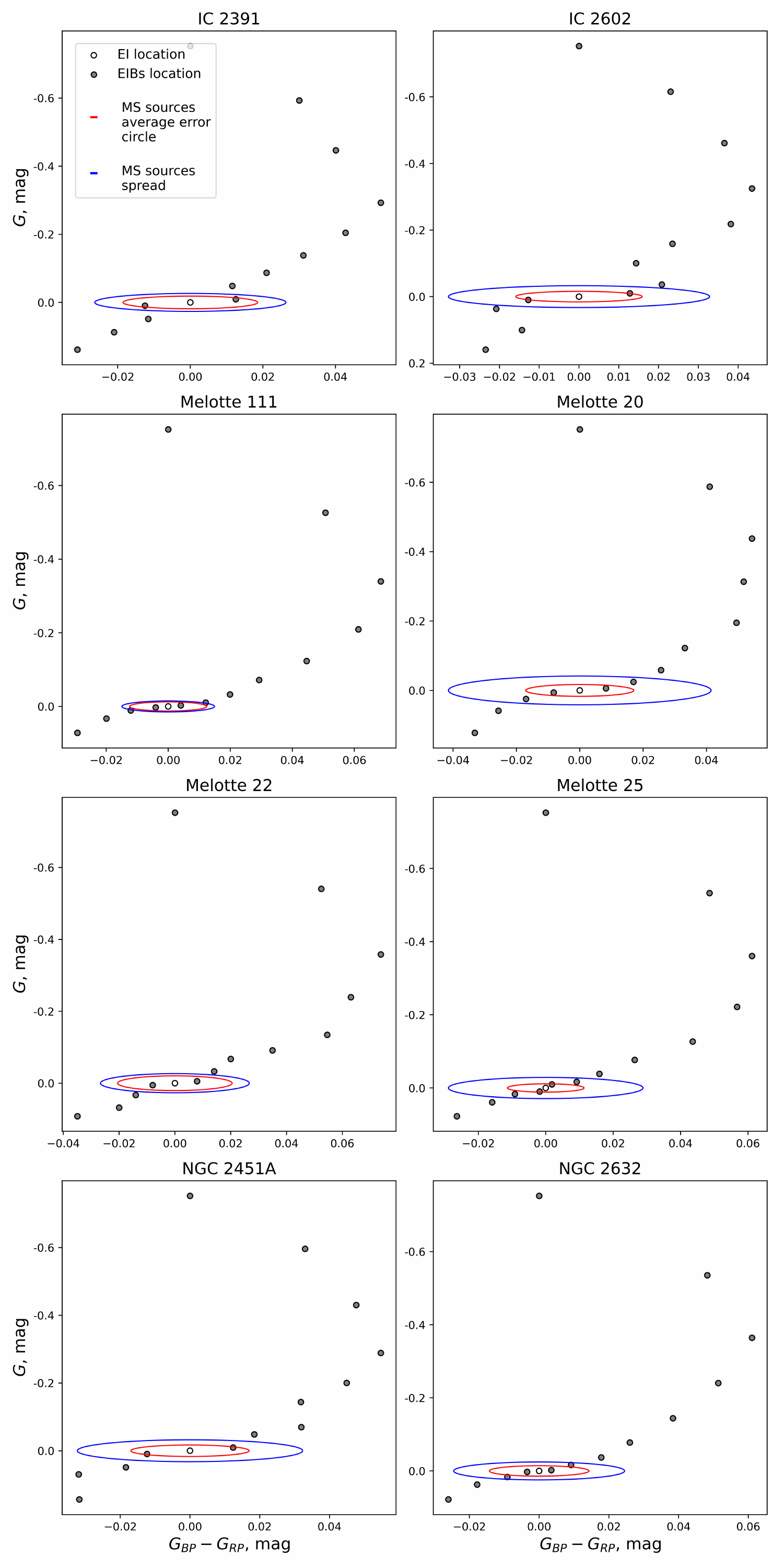}
    \includegraphics[width=0.45\textwidth]{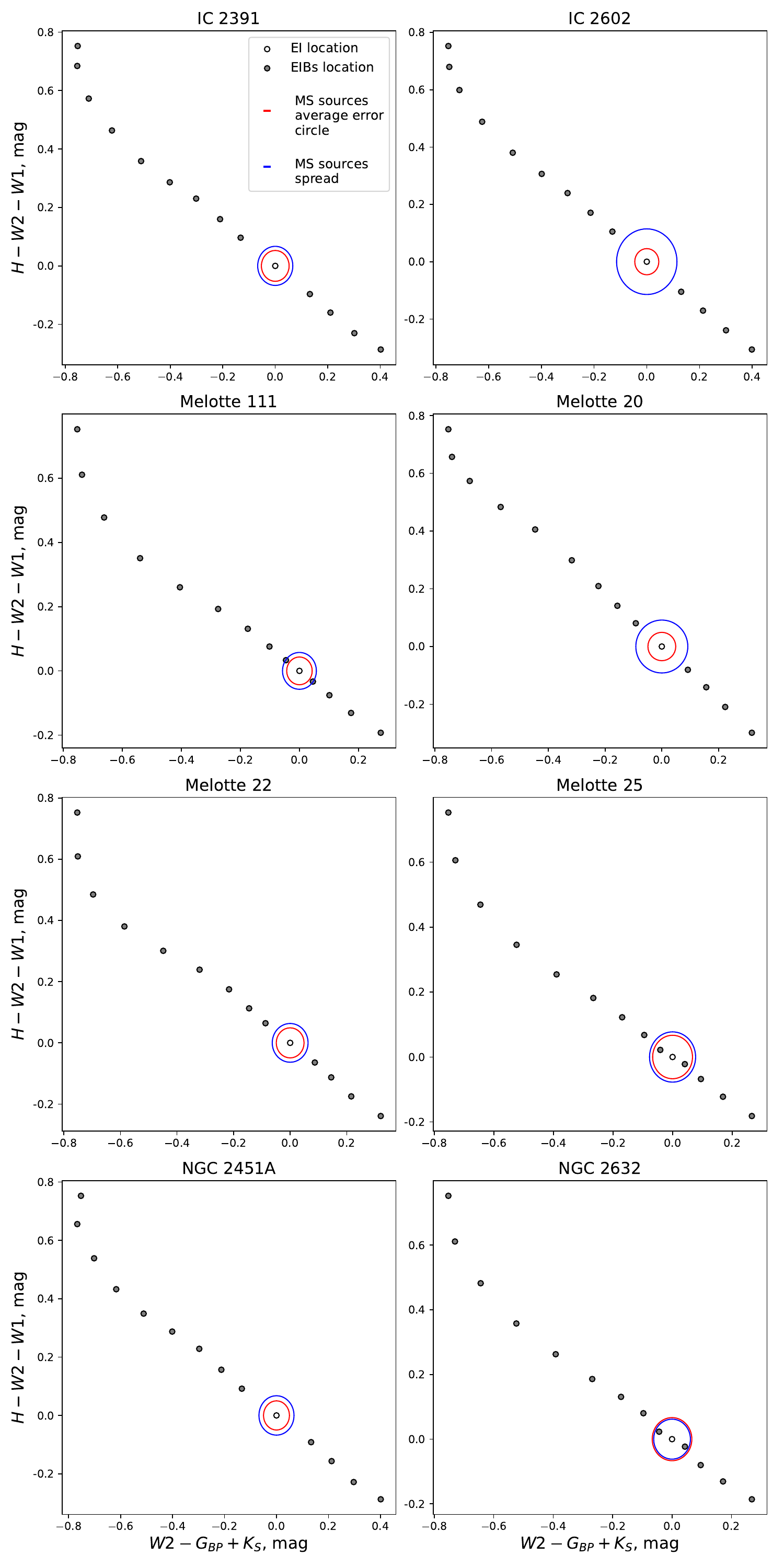}
    \caption{An illustration of CMDs and TIDs with relative location of empirical isochrones of corresponding $q$. Average photometric error of single MS stars (assigned by HDBSCAN) is denoted with red color. An average distance from single stars to EIS is denoted with blue one.}
    \label{appendix_fig:ms_err}
\end{figure}

\newpage
\section{Mass Ratio Distributions}\label{appendix:qDistributions}

\begin{figure}[h!]
    \centering
    \includegraphics[width=0.45\textwidth]{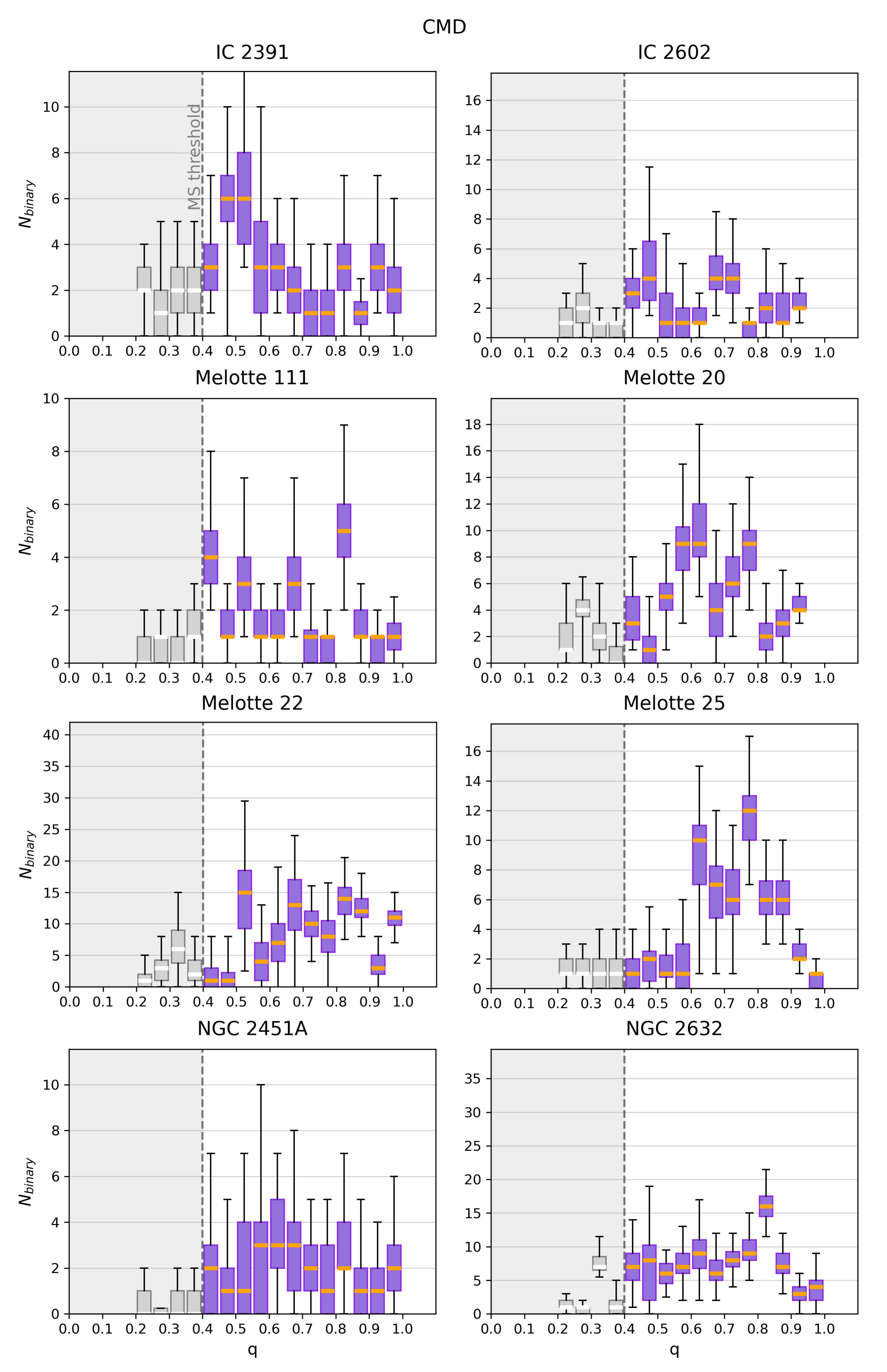}
    \includegraphics[width=0.45\textwidth]{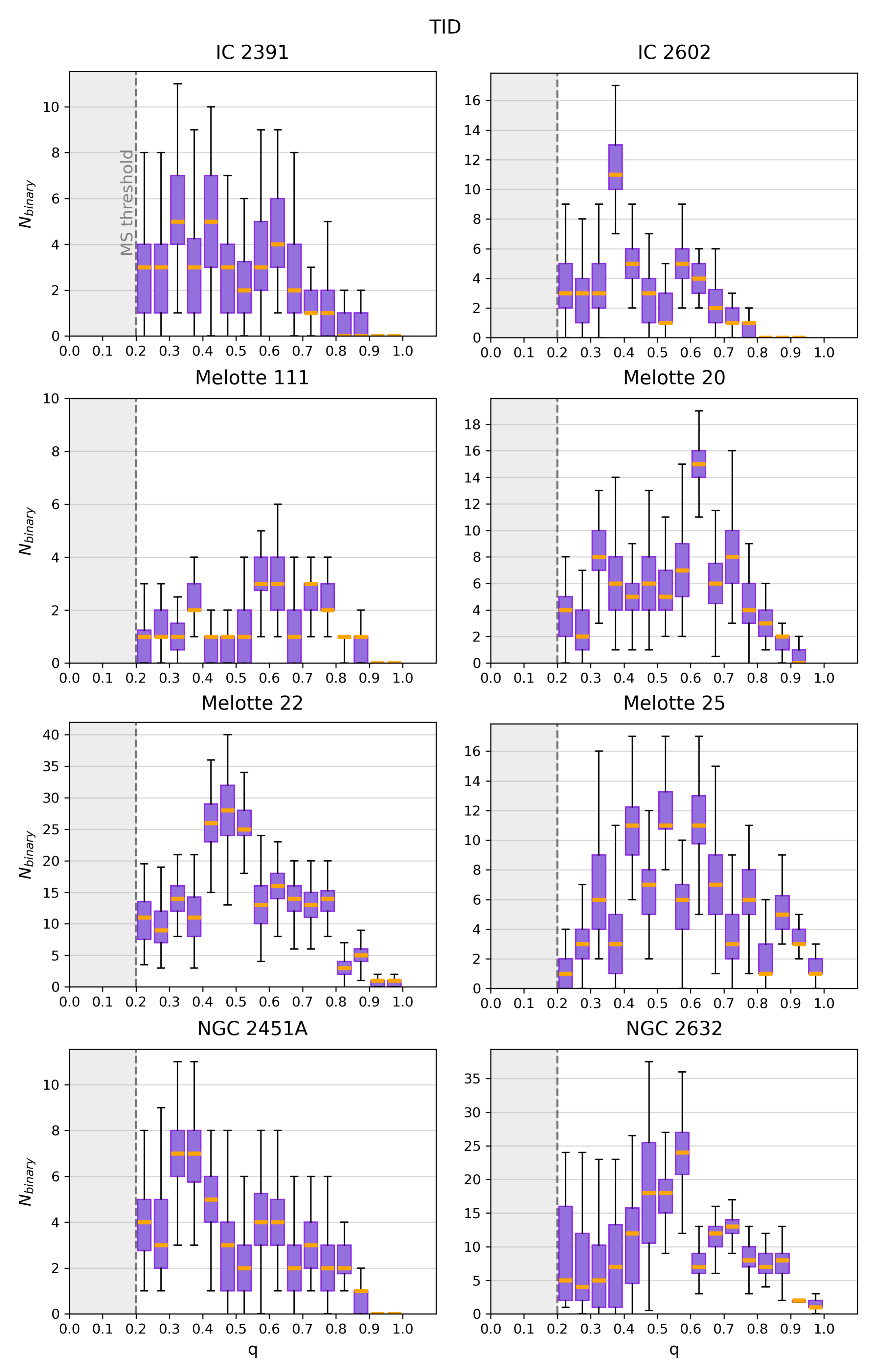}
    \caption{Discrete distribution of the binary mass ratio $q$ for eight open clusters obtained from CMDs and TIDs. The distributions are presented as box-and-whisker plots across fixed $q$-intervals, representing the statistics of 100 Monte Carlo realizations. For each bin, the orange horizontal line indicates the median, the violet box represents the interquartile range (25th–75th percentiles), and the black whiskers denote the range of non-outlier data points. The shaded grey region to the left of the dashed line at $q<0.2$ (for TID-based results) and at $q<0.4$ (for CMD-based results) marks the reliability threshold, below which results are influenced by main sequence contamination and photometric errors.}
    \label{appendix_fig:qDistr_Whisker}
\end{figure}

\bibliography{biblio}{}

@ARTICLE{Griggio2023MNRAS,
       author = {{Griggio}, M. and {Salaris}, M. and {Bedin}, L.~R. and {Cassisi}, S.},
        title = "{The broadening of the main sequence in the open cluster M38}",
      journal = {\mnras},
         year = 2023,
        month = aug,
       volume = {523},
       number = {4},
        pages = {5148-5154},
          doi = {10.1093/mnras/stad1754},
archivePrefix = {arXiv},
       eprint = {2306.05737},
 primaryClass = {astro-ph.SR},
       adsurl = {https://ui.adsabs.harvard.edu/abs/2023MNRAS.523.5148G}
}

@ARTICLE{Carraro2011MNRAS,
       author = {{Carraro}, Giovanni and {Seleznev}, Anton F.},
        title = "{An analysis of the blue straggler population in the Sgr dSph globular cluster Arp 2}",
      journal = {\mnras},
         year = 2011,
        month = apr,
       volume = {412},
       number = {2},
        pages = {1361-1366},
          doi = {10.1111/j.1365-2966.2010.17996.x},
archivePrefix = {arXiv},
       eprint = {1011.1844},
 primaryClass = {astro-ph.GA},
       adsurl = {https://ui.adsabs.harvard.edu/abs/2011MNRAS.412.1361C}
}

@ARTICLE{Kroupa&Jerabkova2018,
       author = {{Kroupa}, Pavel and {Jerabkova}, Tereza},
        title = "{The Impact of Binaries on the Stellar Initial Mass Function}",
      journal = {arXiv e-prints},
         year = 2018,
        month = jun,
          eid = {arXiv:1806.10605},
        pages = {arXiv:1806.10605},
archivePrefix = {arXiv},
       eprint = {1806.10605},
 primaryClass = {astro-ph.GA},
       adsurl = {https://ui.adsabs.harvard.edu/abs/2018arXiv180610605K}
}

@ARTICLE{Kroupa2024arXiv,
       author = {{Kroupa}, Pavel and {Gjergo}, Eda and {Jerabkova}, Tereza and {Yan}, Zhiqiang},
        title = "{The initial mass function of stars}",
      journal = {arXiv e-prints},
         year = 2024,
        month = oct,
          eid = {arXiv:2410.07311},
        pages = {arXiv:2410.07311},
          doi = {10.48550/arXiv.2410.07311},
archivePrefix = {arXiv},
       eprint = {2410.07311},
 primaryClass = {astro-ph.GA},
       adsurl = {https://ui.adsabs.harvard.edu/abs/2024arXiv241007311K}
}

@ARTICLE{Kroupa2001MNRAS,
    author = {{Kroupa}, Pavel},
    title = "{On the variation of the initial mass function}",
    journal = {\mnras},
    year = 2001,
    month = apr,
    volume = {322},
    number = {2},
    pages = {231-246},
    doi = {10.1046/j.1365-8711.2001.04022.x},
    archivePrefix = {arXiv},
    eprint = {astro-ph/0009005},
    primaryClass = {astro-ph},
    adsurl = {https://ui.adsabs.harvard.edu/abs/2001MNRAS.322..231K},
	language = {russian}
}

@ARTICLE{Jiang2024ApJ,
       author = {{Jiang}, Yueyue and {Zhong}, Jing and {Qin}, Songmei and {Tang}, Tong and {Chen}, Li and {Hou}, Jinliang},
        title = "{On the Determination of Stellar Mass and Binary Fraction of Open Clusters within 500 pc from the Sun}",
      journal = {\apj},
         year = 2024,
        month = aug,
       volume = {971},
       number = {1},
          eid = {71},
        pages = {71},
          doi = {10.3847/1538-4357/ad5344},
archivePrefix = {arXiv},
       eprint = {2405.11853},
 primaryClass = {astro-ph.SR},
       adsurl = {https://ui.adsabs.harvard.edu/abs/2024ApJ...971...71J}
}

@ARTICLE{Fisher2005MNRAS,
       author = {{Fisher}, James and {Schr{\"o}der}, Klaus-Peter and {Smith}, Robert Connon},
        title = "{What a local sample of spectroscopic binaries can tell us about the field binary population}",
      journal = {\mnras},
         year = 2005,
        month = aug,
       volume = {361},
       number = {2},
        pages = {495-503},
          doi = {10.1111/j.1365-2966.2005.09193.x},
archivePrefix = {arXiv},
       eprint = {astro-ph/0508651},
 primaryClass = {astro-ph},
       adsurl = {https://ui.adsabs.harvard.edu/abs/2005MNRAS.361..495F}
}

@ARTICLE{Alexander2025MNRAS,
       author = {{Alexander}, Jason S. and {Albrow}, Michael D.},
        title = "{The frequency and mass-ratio distribution of binaries in clusters - III. Probabilistic generative modelling of six young open clusters}",
      journal = {\mnras},
         year = 2025,
        month = jan,
       volume = {536},
       number = {1},
        pages = {471-491},
          doi = {10.1093/mnras/stae2636},
archivePrefix = {arXiv},
       eprint = {2411.16089},
 primaryClass = {astro-ph.GA},
       adsurl = {https://ui.adsabs.harvard.edu/abs/2025MNRAS.536..471A}
}

@ARTICLE{Cordoni2023A&A,
       author = {{Cordoni}, Giacomo and {Milone}, Antonino P. and {Marino}, Anna F. and {Vesperini}, Enrico and {Dondoglio}, Emanuele and {Legnardi}, Maria Vittoria and {Mohandasan}, Anjana and {Carlos}, Marilia and {Lagioia}, Edoardo P. and {Jang}, Sohee and {Ziliotto}, Tuila},
        title = "{Photometric binaries, mass functions, and structural parameters of 78 Galactic open clusters}",
      journal = {\aap},
         year = 2023,
        month = apr,
       volume = {672},
          eid = {A29},
        pages = {A29},
          doi = {10.1051/0004-6361/202245457},
archivePrefix = {arXiv},
       eprint = {2302.03685},
 primaryClass = {astro-ph.SR},
       adsurl = {https://ui.adsabs.harvard.edu/abs/2023A&A...672A..29C}
}

@ARTICLE{Donada2023A&A,
       author = {{Donada}, J. and {Anders}, F. and {Jordi}, C. and {Masana}, E. and {Gieles}, M. and {Perren}, G.~I. and {Balaguer-N{\'u}{\~n}ez}, L. and {Castro-Ginard}, A. and {Cantat-Gaudin}, T. and {Casamiquela}, L.},
        title = "{The multiplicity fraction in 202 open clusters from Gaia}",
      journal = {\aap},
         year = 2023,
        month = jul,
       volume = {675},
          eid = {A89},
        pages = {A89},
          doi = {10.1051/0004-6361/202245219},
archivePrefix = {arXiv},
       eprint = {2301.11061},
 primaryClass = {astro-ph.GA},
       adsurl = {https://ui.adsabs.harvard.edu/abs/2023A&A...675A..89D}
}

@ARTICLE{Muratore2024A&A,
       author = {{Muratore}, F. and {Milone}, A.~P. and {D'Antona}, F. and {Nastasio}, E.~J. and {Cordoni}, G. and {Legnardi}, M.~V. and {He}, C. and {Ziliotto}, T. and {Dondoglio}, E. and {Bernizzoni}, M. and {Tailo}, M. and {Bortolan}, E. and {Dell'Agli}, F. and {Deng}, L. and {Lagioia}, E.~P. and {Li}, C. and {Marino}, A.~F. and {Ventura}, P.},
        title = "{Hubble Space Telescope survey of Magellanic Cloud star clusters. Binaries among the split main sequences of NGC 1818, NGC 1850, and NGC 2164}",
      journal = {\aap},
         year = 2024,
        month = dec,
       volume = {692},
          eid = {A135},
        pages = {A135},
          doi = {10.1051/0004-6361/202451310},
archivePrefix = {arXiv},
       eprint = {2411.02508},
 primaryClass = {astro-ph.SR},
       adsurl = {https://ui.adsabs.harvard.edu/abs/2024A&A...692A.135M}
}

@ARTICLE{Childs2024ApJ,
       author = {{Childs}, Anna C. and {Geller}, Aaron M. and {von Hippel}, Ted and {Motherway}, Erin and {Zwicker}, Claire},
        title = "{Goodbye to Chi by Eye: A Bayesian Analysis of Photometric Binaries in Six Open Clusters}",
      journal = {\apj},
         year = 2024,
        month = feb,
       volume = {962},
       number = {1},
          eid = {41},
        pages = {41},
          doi = {10.3847/1538-4357/ad18c0},
archivePrefix = {arXiv},
       eprint = {2308.16282},
 primaryClass = {astro-ph.SR},
       adsurl = {https://ui.adsabs.harvard.edu/abs/2024ApJ...962...41C}
}

@ARTICLE{Motherway2024ApJ,
       author = {{Motherway}, Erin and {Geller}, Aaron M. and {Childs}, Anna C. and {Zwicker}, Claire and {von Hippel}, Ted},
        title = "{Tracing the Origins of Mass Segregation in M35: Evidence for Primordially Segregated Binaries}",
      journal = {\apjl},
         year = 2024,
        month = feb,
       volume = {962},
       number = {1},
          eid = {L9},
        pages = {L9},
          doi = {10.3847/2041-8213/ad18bf},
archivePrefix = {arXiv},
       eprint = {2308.13520},
 primaryClass = {astro-ph.SR},
       adsurl = {https://ui.adsabs.harvard.edu/abs/2024ApJ...962L...9M}
}

@ARTICLE{He2022ApJ,
       author = {{He}, Chenyu and {Sun}, Weijia and {Li}, Chengyuan and {Li}, Lu and {Shao}, Zhengyi and {Zhong}, Jing and {Chen}, Li and {de Grijs}, Richard and {Tang}, Baitian and {Qin}, Songmei and {Randriamanakoto}, Zara},
        title = "{The Role of Binarity and Stellar Rotation in the Split Main Sequence of NGC 2422}",
      journal = {\apj},
         year = 2022,
        month = oct,
       volume = {938},
       number = {1},
          eid = {42},
        pages = {42},
          doi = {10.3847/1538-4357/ac8b08},
archivePrefix = {arXiv},
       eprint = {2208.10843},
 primaryClass = {astro-ph.SR},
       adsurl = {https://ui.adsabs.harvard.edu/abs/2022ApJ...938...42H}
}

@ARTICLE{Chulkov2025AJ,
       author = {{Chulkov}, Dmitry and {Strakhov}, Ivan and {Safonov}, Boris},
        title = "{Resolving Pleiades Binary Stars with Gaia and Speckle Interferometric Observations}",
      journal = {\aj},
         year = 2025,
        month = mar,
       volume = {169},
       number = {3},
          eid = {145},
        pages = {145},
          doi = {10.3847/1538-3881/ada564},
archivePrefix = {arXiv},
       eprint = {2412.20986},
 primaryClass = {astro-ph.SR},
       adsurl = {https://ui.adsabs.harvard.edu/abs/2025AJ....169..145C}
}

@ARTICLE{Chulkov2024AJ,
       author = {{Chulkov}, Dmitry},
        title = "{Unveiling Subarcsecond Multiplicity in the Pleiades with Gaia Multicolor Photometry}",
      journal = {\aj},
         year = 2024,
        month = oct,
       volume = {168},
       number = {4},
          eid = {156},
        pages = {156},
          doi = {10.3847/1538-3881/ad7025},
archivePrefix = {arXiv},
       eprint = {2408.05423},
 primaryClass = {astro-ph.SR},
       adsurl = {https://ui.adsabs.harvard.edu/abs/2024AJ....168..156C}
}

@ARTICLE{Zwicker2024ApJ,
       author = {{Zwicker}, Claire and {Geller}, Aaron M. and {Childs}, Anna C. and {Motherway}, Erin and {von Hippel}, Ted},
        title = "{Investigating Mass Segregation of the Binary Stars in the Open Cluster NGC 6819}",
      journal = {\apj},
         year = 2024,
        month = may,
       volume = {967},
       number = {1},
          eid = {44},
        pages = {44},
          doi = {10.3847/1538-4357/ad39c6},
archivePrefix = {arXiv},
       eprint = {2308.15582},
 primaryClass = {astro-ph.SR},
       adsurl = {https://ui.adsabs.harvard.edu/abs/2024ApJ...967...44Z}
}

@ARTICLE{Spangler2025RNAAS,
       author = {{Spangler}, Steven R.},
        title = "{Broadening of the Main Sequence of a Star Cluster by Undetected Binaries}",
      journal = {Research Notes of the American Astronomical Society},
         year = 2025,
        month = feb,
       volume = {9},
       number = {2},
          eid = {34},
        pages = {34},
          doi = {10.3847/2515-5172/adb48a},
archivePrefix = {arXiv},
       eprint = {2501.17306},
 primaryClass = {astro-ph.SR},
       adsurl = {https://ui.adsabs.harvard.edu/abs/2025RNAAS...9...34S}
}

@ARTICLE{Liu2025AJ,
       author = {{Liu}, Rongrong and {Shao}, Zhengyi and {Li}, Lu},
        title = "{Photometric Determination of Unresolved Main-sequence Binaries in the Pleiades: Binary Fraction and Mass-ratio Distribution}",
      journal = {\aj},
         year = 2025,
        month = feb,
       volume = {169},
       number = {2},
          eid = {116},
        pages = {116},
          doi = {10.3847/1538-3881/ada380},
archivePrefix = {arXiv},
       eprint = {2501.01617},
 primaryClass = {astro-ph.SR},
       adsurl = {https://ui.adsabs.harvard.edu/abs/2025AJ....169..116L}
}

@ARTICLE{Liu2023ApJS,
       author = {{Long}, Liu and {Bi}, Shaolan and {Zhang}, Jinghua and {Zhang}, Xianfei and {Zhang}, Liyun and {Ge}, Zhishuai and {Li}, Tanda and {Chen}, Xunzhou and {Li}, YaGuang and {Ye}, LiFei and {Sun}, TianCheng and {Zhou}, JianZhao},
        title = "{Investigating 16 Open Clusters in the Kepler/K2-Gaia DR3 Field. I. Membership, Binary Systems, and Rotation}",
      journal = {\apjs},
         year = 2023,
        month = sep,
       volume = {268},
       number = {1},
          eid = {30},
        pages = {30},
          doi = {10.3847/1538-4365/ace5af},
archivePrefix = {arXiv},
       eprint = {2307.06596},
 primaryClass = {astro-ph.GA},
       adsurl = {https://ui.adsabs.harvard.edu/abs/2023ApJS..268...30L}
}

@ARTICLE{Jadhav2021AJ,
       author = {{Jadhav}, Vikrant V. and {Roy}, Kaustubh and {Joshi}, Naman and {Subramaniam}, Annapurni},
        title = "{High Mass-Ratio Binary Population in Open Clusters: Segregation of Early Type Binaries and an Increasing Binary Fraction with Mass}",
      journal = {\aj},
         year = 2021,
        month = dec,
       volume = {162},
       number = {6},
          eid = {264},
        pages = {264},
          doi = {10.3847/1538-3881/ac2571},
archivePrefix = {arXiv},
       eprint = {2109.03782},
 primaryClass = {astro-ph.SR},
       adsurl = {https://ui.adsabs.harvard.edu/abs/2021AJ....162..264J}
}

@ARTICLE{Wang2023ApJ,
       author = {{Wang}, Li and {Li}, Chengyuan and {Wang}, Long and {He}, Chenyu and {Wang}, Chen},
        title = "{On the Origin of the Split Main Sequences of the Young Massive Cluster NGC 1856}",
      journal = {\apj},
         year = 2023,
        month = jun,
       volume = {949},
       number = {2},
          eid = {53},
        pages = {53},
          doi = {10.3847/1538-4357/accae0},
archivePrefix = {arXiv},
       eprint = {2304.02227},
 primaryClass = {astro-ph.SR},
       adsurl = {https://ui.adsabs.harvard.edu/abs/2023ApJ...949...53W}
}

@ARTICLE{Chengyuan2024arXiv,
       author = {{Li}, Chengyuan and {Milone}, Antonino P. and {Sun}, Weijia and {de Grijs}, Richard},
        title = "{Extended Main Sequences in Star Clusters}",
      journal = {arXiv e-prints},
         year = 2024,
        month = jan,
          eid = {arXiv:2401.08062},
        pages = {arXiv:2401.08062},
          doi = {10.48550/arXiv.2401.08062},
archivePrefix = {arXiv},
       eprint = {2401.08062},
 primaryClass = {astro-ph.SR},
       adsurl = {https://ui.adsabs.harvard.edu/abs/2024arXiv240108062L}
}

@ARTICLE{Niu2020ApJ,
       author = {{Niu}, Hubiao and {Wang}, Jiaxin and {Fu}, Jianning},
        title = "{Binary Fraction Estimation of Main-sequence Stars in 12 Open Clusters: Based on the Homogeneous Data of LAMOST Survey and Gaia DR2}",
      journal = {\apj},
         year = 2020,
        month = nov,
       volume = {903},
       number = {2},
          eid = {93},
        pages = {93},
          doi = {10.3847/1538-4357/abb8d6},
       adsurl = {https://ui.adsabs.harvard.edu/abs/2020ApJ...903...93N}
}

@ARTICLE{Kouwenhoven+2009,
       author = {{Kouwenhoven}, M.~B.~N. and {Brown}, A.~G.~A. and {Goodwin}, S.~P. and {Portegies Zwart}, S.~F. and {Kaper}, L.},
        title = "{Exploring the consequences of pairing algorithms for binary stars}",
      journal = {\aap},
         year = 2009,
        month = jan,
       volume = {493},
       number = {3},
        pages = {979-1016},
          doi = {10.1051/0004-6361:200810234},
archivePrefix = {arXiv},
       eprint = {0811.2859},
 primaryClass = {astro-ph},
       adsurl = {https://ui.adsabs.harvard.edu/abs/2009A&A...493..979K}
}

@ARTICLE{Bressan+2012,
       author = {{Bressan}, Alessandro and {Marigo}, Paola and {Girardi}, L{\'e}o. and {Salasnich}, Bernardo and {Dal Cero}, Claudia and {Rubele}, Stefano and {Nanni}, Ambra},
        title = "{PARSEC: stellar tracks and isochrones with the PAdova and TRieste Stellar Evolution Code}",
      journal = {\mnras},
         year = 2012,
        month = nov,
       volume = {427},
       number = {1},
        pages = {127-145},
          doi = {10.1111/j.1365-2966.2012.21948.x},
archivePrefix = {arXiv},
       eprint = {1208.4498},
 primaryClass = {astro-ph.SR},
       adsurl = {https://ui.adsabs.harvard.edu/abs/2012MNRAS.427..127B}
}

@ARTICLE{Danilov&Seleznev2020,
       author = {{Danilov}, V.~M. and {Seleznev}, A.~F.},
        title = "{On the Motion of Stars in the Pleiades According to Gaia DR2 Data}",
      journal = {Astrophysical Bulletin},
         year = 2020,
        month = oct,
       volume = {75},
       number = {4},
        pages = {407-424},
          doi = {10.1134/S1990341320040045},
archivePrefix = {arXiv},
       eprint = {2012.15289},
 primaryClass = {astro-ph.GA},
       adsurl = {https://ui.adsabs.harvard.edu/abs/2020AstBu..75..407D}
}

@ARTICLE{Lodieu+2019,
       author = {{Lodieu}, N. and {P{\'e}rez-Garrido}, A. and {Smart}, R.~L. and {Silvotti}, R.},
        title = "{A 5D view of the {\ensuremath{\alpha}} Per, Pleiades, and Praesepe clusters}",
      journal = {\aap},
         year = 2019,
        month = aug,
       volume = {628},
          eid = {A66},
        pages = {A66},
          doi = {10.1051/0004-6361/201935533},
archivePrefix = {arXiv},
       eprint = {1906.03924},
 primaryClass = {astro-ph.SR},
       adsurl = {https://ui.adsabs.harvard.edu/abs/2019A&A...628A..66L}
}

@ARTICLE{Dias+2021,
       author = {{Dias}, W.~S. and {Monteiro}, H. and {Moitinho}, A. and {L{\'e}pine}, J.~R.~D. and {Carraro}, G. and {Paunzen}, E. and {Alessi}, B. and {Villela}, L.},
        title = "{Updated parameters of 1743 open clusters based on Gaia DR2}",
      journal = {\mnras},
         year = 2021,
        month = jun,
       volume = {504},
       number = {1},
        pages = {356-371},
          doi = {10.1093/mnras/stab770},
archivePrefix = {arXiv},
       eprint = {2103.12829},
 primaryClass = {astro-ph.SR},
       adsurl = {https://ui.adsabs.harvard.edu/abs/2021MNRAS.504..356D}
}

@ARTICLE{Malofeeva+2022,
       author = {{Malofeeva}, Alina A. and {Seleznev}, Anton F. and {Carraro}, Giovanni},
        title = "{Unresolved Binaries in the Intermediate Mass Range in the Pleiades Star Cluster}",
      journal = {\aj},
         year = 2022,
        month = mar,
       volume = {163},
       number = {3},
          eid = {113},
        pages = {113},
          doi = {10.3847/1538-3881/ac47a3},
archivePrefix = {arXiv},
       eprint = {2201.05146},
 primaryClass = {astro-ph.SR},
       adsurl = {https://ui.adsabs.harvard.edu/abs/2022AJ....163..113M}
}

@ARTICLE{Malofeeva+2023,
       author = {{Malofeeva}, Alina A. and {Mikhnevich}, Varvara O. and {Carraro}, Giovanni and {Seleznev}, Anton F.},
        title = "{Unresolved Binaries and Multiples in the Intermediate Mass Range in Open Clusters: Pleiades, Alpha Per, Praesepe, and NGC 1039}",
      journal = {\aj},
         year = 2023,
        month = feb,
       volume = {165},
       number = {2},
          eid = {45},
        pages = {45},
          doi = {10.3847/1538-3881/aca666},
archivePrefix = {arXiv},
       eprint = {2211.12745},
 primaryClass = {astro-ph.GA},
       adsurl = {https://ui.adsabs.harvard.edu/abs/2023AJ....165...45M}
}

@INPROCEEDINGS{Mikhnevich2024maeu.conf,
       author = {{Mikhnevich}, V. and {Plotnikova}, A. and {Seleznev}, A. and {Carraro}, G.},
        title = "{Empirical isochrones for single and binary stars in open clusters}",
    booktitle = {Modern Astronomy: From the Early Universe to Exoplanets and Black Holes (VAK2024},
         year = 2024,
        month = dec,
        pages = {443-449},
          doi = {10.26119/VAK2024.070},
       adsurl = {https://ui.adsabs.harvard.edu/abs/2024maeu.conf..443M}
}

@ARTICLE{Nikiforova+2020,
       author = {{Nikiforova}, Victoria V. and {Kulesh}, Maxim V. and {Seleznev}, Anton F. and {Carraro}, Giovanni},
        title = "{The Relation of the Alpha Persei Star Cluster with the Nearby Stellar Stream}",
      journal = {\aj},
         year = 2020,
        month = sep,
       volume = {160},
       number = {3},
          eid = {142},
        pages = {142},
          doi = {10.3847/1538-3881/aba753},
archivePrefix = {arXiv},
       eprint = {2007.11211},
 primaryClass = {astro-ph.GA},
       adsurl = {https://ui.adsabs.harvard.edu/abs/2020AJ....160..142N}
}

@ARTICLE{Cantat-Gaudin+2020,
       author = {{Cantat-Gaudin}, T. and {Anders}, F. and {Castro-Ginard}, A. and {Jordi}, C. and {Romero-G{\'o}mez}, M. and {Soubiran}, C. and {Casamiquela}, L. and {Tarricq}, Y. and {Moitinho}, A. and {Vallenari}, A. and {Bragaglia}, A. and {Krone-Martins}, A. and {Kounkel}, M.},
        title = "{Painting a portrait of the Galactic disc with its stellar clusters}",
      journal = {\aap},
         year = 2020,
        month = aug,
       volume = {640},
          eid = {A1},
        pages = {A1},
          doi = {10.1051/0004-6361/202038192},
archivePrefix = {arXiv},
       eprint = {2004.07274},
 primaryClass = {astro-ph.GA},
       adsurl = {https://ui.adsabs.harvard.edu/abs/2020A&A...640A...1C}
}

@INPROCEEDINGS{Malkov+2010,
       author = {{Malkov}, O. and {Mironov}, A. and {Sichevskij}, S.},
        title = "{Recognition of unresolved binaries on Gaia colour index diagrams}",
    booktitle = {EAS Publications Series},
         year = 2010,
       series = {EAS Publications Series},
       volume = {45},
        month = jan,
        pages = {409-412},
          doi = {10.1051/eas/1045071},
       adsurl = {https://ui.adsabs.harvard.edu/abs/2010EAS....45..409M}
}

@ARTICLE{Malkov+2011,
       author = {{Malkov}, Oleg and {Mironov}, Aleksej and {Sichevskij}, Sergej},
        title = "{Single-binary star separation by ultraviolet color index diagrams}",
      journal = {\apss},
         year = 2011,
        month = sep,
       volume = {335},
       number = {1},
        pages = {105-111},
          doi = {10.1007/s10509-011-0613-1},
       adsurl = {https://ui.adsabs.harvard.edu/abs/2011Ap&SS.335..105M}
}

@ARTICLE{El-Badry+2019,
       author = {{El-Badry}, Kareem and {Rix}, Hans-Walter and {Tian}, Haijun and {Duch{\^e}ne}, Gaspard and {Moe}, Maxwell},
        title = "{Discovery of an equal-mass `twin' binary population reaching 1000 + au separations}",
      journal = {\mnras},
         year = 2019,
        month = nov,
       volume = {489},
       number = {4},
        pages = {5822-5857},
          doi = {10.1093/mnras/stz2480},
archivePrefix = {arXiv},
       eprint = {1906.10128},
 primaryClass = {astro-ph.SR},
       adsurl = {https://ui.adsabs.harvard.edu/abs/2019MNRAS.489.5822E}
}

@ARTICLE{Mikhnevich&Seleznev,
       author = {{Mikhnevich}, V.~O. and {Seleznev}, A.~F.},
        title = "{Unresolved Binary Systems with White Dwarfs in Open Star Clusters}",
      journal = {Astronomy Reports},
         year = 2024,
        month = feb,
       volume = {68},
       number = {2},
        pages = {121-128},
          doi = {10.1134/S1063772924700161},
       adsurl = {https://ui.adsabs.harvard.edu/abs/2024ARep...68..121M}
}

@ARTICLE{Hunt&Reffert2024,
       author = {{Hunt}, Emily L. and {Reffert}, Sabine},
        title = "{Improving the open cluster census. III. Using cluster masses, radii, and dynamics to create a cleaned open cluster catalogue}",
      journal = {\aap},
         year = 2024,
        month = jun,
       volume = {686},
          eid = {A42},
        pages = {A42},
          doi = {10.1051/0004-6361/202348662},
archivePrefix = {arXiv},
       eprint = {2403.05143},
 primaryClass = {astro-ph.GA},
       adsurl = {https://ui.adsabs.harvard.edu/abs/2024A&A...686A..42H}
}

@ARTICLE{Angelo2023MNRAS,
       author = {{Angelo}, M.~S. and {Santos}, Jr., J.~F.~C. and {Maia}, F.~F.~S. and {Corradi}, W.~J.~B.},
        title = "{Enlightening the dynamical evolution of Galactic open clusters: an approach using Gaia DR3 and analytical descriptions}",
      journal = {\mnras},
         year = 2023,
        month = jun,
       volume = {522},
       number = {1},
        pages = {956-975},
          doi = {10.1093/mnras/stad1038},
archivePrefix = {arXiv},
       eprint = {2304.02209},
 primaryClass = {astro-ph.GA},
       adsurl = {https://ui.adsabs.harvard.edu/abs/2023MNRAS.522..956A}
}

@ARTICLE{Angelo2025arXiv,
       author = {{Angelo}, M.~S. and {Santos}, Jr., J.~F.~C. and {Corradi}, W.~J.~B. and {Maia}, F.~F.~S.},
        title = "{Exploring the dynamical state of Galactic open clusters using Gaia DR3 and observational parameters}",
      journal = {arXiv e-prints},
         year = 2025,
        month = apr,
          eid = {arXiv:2504.06362},
        pages = {arXiv:2504.06362},
          doi = {10.48550/arXiv.2504.06362},
archivePrefix = {arXiv},
       eprint = {2504.06362},
 primaryClass = {astro-ph.GA},
       adsurl = {https://ui.adsabs.harvard.edu/abs/2025arXiv250406362A}
}

@ARTICLE{Tagaev2025,
        author    = "{Tagaev}, D. I. and {Seleznev}, A. F.",
        title     = "Search for the possible members of the open cluster {NGC} 3532 with poor astrometric solutions of {GAIA} {DR3}",
        journal   = "{Astronomy Reports}",
        publisher = "{Pleiades Publishing Ltd}",
        volume    =  {69},
        number    =  {6},
        pages     = {457-468},
        month     =  jun,
        year      =  2025,
        doi = {10.1134/S1063772925701884},
        copyright = "https://www.springernature.com/gp/researchers/text-and-data-mining"
}

@ARTICLE{astroquery,
   author = {{Ginsburg}, A. and {Sip{\H o}cz}, B.~M. and {Brasseur}, C.~E. and
	{Cowperthwaite}, P.~S. and {Craig}, M.~W. and {Deil}, C. and
	{Guillochon}, J. and {Guzman}, G. and {Liedtke}, S. and {Lian Lim}, P. and
	{Lockhart}, K.~E. and {Mommert}, M. and {Morris}, B.~M. and
	{Norman}, H. and {Parikh}, M. and {Persson}, M.~V. and {Robitaille}, T.~P. and
	{Segovia}, J.-C. and {Singer}, L.~P. and {Tollerud}, E.~J. and
	{de Val-Borro}, M. and {Valtchanov}, I. and {Woillez}, J. and
	{The Astroquery collaboration} and {a subset of the astropy collaboration}
	},
    title = "{astroquery: An Astronomical Web-querying Package in Python}",
  journal = {\aj},
archivePrefix = "arXiv",
   eprint = {1901.04520},
 primaryClass = "astro-ph.IM",
     year = 2019,
    month = mar,
   volume = 157,
      eid = {98},
    pages = {98},
      doi = {10.3847/1538-3881/aafc33},
   adsurl = {https://adsabs.harvard.edu/abs/2019AJ....157...98G}
}

@article{McInnes2017,
  doi = {10.21105/joss.00205},
  url = {https://doi.org/10.21105%2Fjoss.00205},
  year  = {2017},
  month = {mar},
  publisher = {The Open Journal},
  volume = {2},
  number = {11},
  author = {Leland McInnes and John Healy and Steve Astels},
  title = {hdbscan: Hierarchical density based clustering},
  journal = {The Journal of Open Source Software}
}

@ARTICLE{McInnes2017arXiv,
       author = {{McInnes}, Leland and {Healy}, John},
        title = "{Accelerated Hierarchical Density Clustering}",
      journal = {arXiv e-prints},
         year = 2017,
        month = may,
          eid = {arXiv:1705.07321},
        pages = {arXiv:1705.07321},
          doi = {10.48550/arXiv.1705.07321},
archivePrefix = {arXiv},
       eprint = {1705.07321},
 primaryClass = {stat.ML},
       adsurl = {https://ui.adsabs.harvard.edu/abs/2017arXiv170507321M}
}

@ARTICLE{Wright2020NewAR,
       author = {{Wright}, Nicholas J.},
        title = "{OB Associations and their origins}",
      journal = {\nar},
         year = 2020,
        month = nov,
       volume = {90},
          eid = {101549},
        pages = {101549},
          doi = {10.1016/j.newar.2020.101549},
archivePrefix = {arXiv},
       eprint = {2011.09483},
 primaryClass = {astro-ph.SR},
       adsurl = {https://ui.adsabs.harvard.edu/abs/2020NewAR..9001549W}
}

@ARTICLE{Haffner&Heckmann1937,
       author = {{Haffner}, Hans and {Heckmann}, Otto},
        title = "{Das Farben-Helligkeits-Diagramm der Praesepe auf Grund neuer Beobachtungen}",
      journal = {Veroeffentlichungen der Universitaets-Sternwarte zu Goettingen},
         year = 1937,
        month = jan,
       volume = {0004},
        pages = {77-95},
       adsurl = {https://ui.adsabs.harvard.edu/abs/1937VeGoe...4...77H}
}

@ARTICLE{Maxted+2008,
       author = {{Maxted}, P.~F.~L. and {Jeffries}, R.~D. and {Oliveira}, J.~M. and {Naylor}, T. and {Jackson}, R.~J.},
        title = "{A survey for low-mass spectroscopic binary stars in the young clusters around {\ensuremath{\sigma}} Orionis and {\ensuremath{\lambda}} Orionis}",
      journal = {\mnras},
         year = 2008,
        month = apr,
       volume = {385},
       number = {4},
        pages = {2210-2224},
          doi = {10.1111/j.1365-2966.2008.13008.x},
archivePrefix = {arXiv},
       eprint = {0801.3595},
 primaryClass = {astro-ph},
       adsurl = {https://ui.adsabs.harvard.edu/abs/2008MNRAS.385.2210M}
}

@ARTICLE{Duquennoy&Mayor1991,
       author = {{Duquennoy}, A. and {Mayor}, M.},
        title = "{Multiplicity among Solar Type Stars in the Solar Neighbourhood - Part Two - Distribution of the Orbital Elements in an Unbiased Sample}",
      journal = {\aap},
         year = 1991,
        month = aug,
       volume = {248},
        pages = {485},
       adsurl = {https://ui.adsabs.harvard.edu/abs/1991A&A...248..485D}
}

@ARTICLE{BardalezGagliuffi+2014,
       author = {{Bardalez Gagliuffi}, Daniella C. and {Burgasser}, Adam J. and {Gelino}, Christopher R. and {Looper}, Dagny L. and {Nicholls}, Christine P. and {Schmidt}, Sarah J. and {Cruz}, Kelle and {West}, Andrew A. and {Gizis}, John E. and {Metchev}, Stanimir},
        title = "{SpeX Spectroscopy of Unresolved Very Low Mass Binaries. II. Identification of 14 Candidate Binaries with Late-M/Early-L and T Dwarf Components}",
      journal = {\apj},
         year = 2014,
        month = oct,
       volume = {794},
       number = {2},
          eid = {143},
        pages = {143},
          doi = {10.1088/0004-637X/794/2/143},
archivePrefix = {arXiv},
       eprint = {1408.3089},
 primaryClass = {astro-ph.SR},
       adsurl = {https://ui.adsabs.harvard.edu/abs/2014ApJ...794..143B}
}

@ARTICLE{Raghavan+2010,
       author = {{Raghavan}, Deepak and {McAlister}, Harold A. and {Henry}, Todd J. and {Latham}, David W. and {Marcy}, Geoffrey W. and {Mason}, Brian D. and {Gies}, Douglas R. and {White}, Russel J. and {ten Brummelaar}, Theo A.},
        title = "{A Survey of Stellar Families: Multiplicity of Solar-type Stars}",
      journal = {\apjs},
         year = 2010,
        month = sep,
       volume = {190},
       number = {1},
        pages = {1-42},
          doi = {10.1088/0067-0049/190/1/1},
archivePrefix = {arXiv},
       eprint = {1007.0414},
 primaryClass = {astro-ph.SR},
       adsurl = {https://ui.adsabs.harvard.edu/abs/2010ApJS..190....1R}
}

@ARTICLE{Bonifazi+1990,
       author = {{Bonifazi}, A. and {Fusi-Pecci}, F. and {Romeo}, G. and {Tosi}, M.},
        title = "{CCD Photometry of Galactic Open Clusters - Part Two - NGC2243}",
      journal = {\mnras},
         year = 1990,
        month = jul,
       volume = {245},
        pages = {15},
          doi = {10.1093/mnras/245.1.15},
       adsurl = {https://ui.adsabs.harvard.edu/abs/1990MNRAS.245...15B}
}

@ARTICLE{Khalaj&Baumgardt2013,
       author = {{Khalaj}, P. and {Baumgardt}, H.},
        title = "{The stellar mass function, binary content and radial structure of the open cluster Praesepe derived from PPMXL and SDSS data}",
      journal = {\mnras},
         year = 2013,
        month = oct,
       volume = {434},
       number = {4},
        pages = {3236-3245},
          doi = {10.1093/mnras/stt1239},
archivePrefix = {arXiv},
       eprint = {1307.2020},
 primaryClass = {astro-ph.GA},
       adsurl = {https://ui.adsabs.harvard.edu/abs/2013MNRAS.434.3236K}
}

@ARTICLE{Sarro+2014,
       author = {{Sarro}, L.~M. and {Bouy}, H. and {Berihuete}, A. and {Bertin}, E. and {Moraux}, E. and {Bouvier}, J. and {Cuillandre}, J. -C. and {Barrado}, D. and {Solano}, E.},
        title = "{Cluster membership probabilities from proper motions and multi-wavelength photometric catalogues. I. Method and application to the Pleiades cluster}",
      journal = {\aap},
         year = 2014,
        month = mar,
       volume = {563},
          eid = {A45},
        pages = {A45},
          doi = {10.1051/0004-6361/201322413},
archivePrefix = {arXiv},
       eprint = {1401.7427},
 primaryClass = {astro-ph.SR},
       adsurl = {https://ui.adsabs.harvard.edu/abs/2014A&A...563A..45S}
}

@ARTICLE{Sheikhi+2016,
       author = {{Sheikhi}, Najmeh and {Hasheminia}, Maryam and {Khalaj}, Pouria and {Haghi}, Hosein and {Zonoozi}, Akram Hasani and {Baumgardt}, Holger},
        title = "{The binary fraction and mass segregation in Alpha Persei open cluster}",
      journal = {\mnras},
         year = 2016,
        month = mar,
       volume = {457},
       number = {1},
        pages = {1028-1036},
          doi = {10.1093/mnras/stw059},
archivePrefix = {arXiv},
       eprint = {1601.02186},
 primaryClass = {astro-ph.GA},
       adsurl = {https://ui.adsabs.harvard.edu/abs/2016MNRAS.457.1028S}
}

@ARTICLE{Jadhav2021MNRAS,
       author = {{Jadhav}, Vikrant V. and {Subramaniam}, Annapurni},
        title = "{Blue straggler stars in open clusters using Gaia: dependence on cluster parameters and possible formation pathways}",
      journal = {\mnras},
         year = 2021,
        month = oct,
       volume = {507},
       number = {2},
        pages = {1699-1709},
          doi = {10.1093/mnras/stab2264},
archivePrefix = {arXiv},
       eprint = {2108.01396},
 primaryClass = {astro-ph.SR},
       adsurl = {https://ui.adsabs.harvard.edu/abs/2021MNRAS.507.1699J}
}

@ARTICLE{Harvey2024RNAAS,
       author = {{Harvey}, Andrew and {Sahu}, Yuvraj and {Flores}, Elijah},
        title = "{An Analysis of Blue Straggler Stars in Open Clusters}",
      journal = {Research Notes of the American Astronomical Society},
         year = 2024,
        month = jul,
       volume = {8},
       number = {7},
          eid = {191},
        pages = {191},
          doi = {10.3847/2515-5172/ad686e},
       adsurl = {https://ui.adsabs.harvard.edu/abs/2024RNAAS...8..191H}
}

@ARTICLE{Liebert1994AJ,
       author = {{Liebert}, James and {Saffer}, Rex A. and {Green}, Elizabeth M.},
        title = "{The Evolved Hot Stars of The Old, Metal-Rich Galactic Cluster NGC 6791}",
      journal = {\aj},
         year = 1994,
        month = apr,
       volume = {107},
        pages = {1408},
          doi = {10.1086/116954},
       adsurl = {https://ui.adsabs.harvard.edu/abs/1994AJ....107.1408L}
}

@ARTICLE{Carraro2017ApJL,
       author = {{Carraro}, Giovanni and {Benvenuto}, Omar G.},
        title = "{Binarity as the Solution to the Stellar Evolution Enigma Posed by NGC 6791}",
      journal = {\apjl},
         year = 2017,
        month = may,
       volume = {841},
       number = {1},
          eid = {L10},
        pages = {L10},
          doi = {10.3847/2041-8213/aa7131},
archivePrefix = {arXiv},
       eprint = {1705.01776},
 primaryClass = {astro-ph.SR},
       adsurl = {https://ui.adsabs.harvard.edu/abs/2017ApJ...841L..10C}
}

@ARTICLE{Childs2025arXiv,
       author = {{Childs}, Anna C. and {Geller}, Aaron M.},
        title = "{Stellar Dynamics in Open Clusters Increases the Binary Fraction and Mass Ratios: Evidence from Photometric Binaries in 35 Open Clusters}",
      journal = {arXiv e-prints},
         year = 2025,
        month = jun,
          eid = {arXiv:2506.20889},
        pages = {arXiv:2506.20889},
          doi = {10.48550/arXiv.2506.20889},
archivePrefix = {arXiv},
       eprint = {2506.20889},
 primaryClass = {astro-ph.SR},
       adsurl = {https://ui.adsabs.harvard.edu/abs/2025arXiv250620889C}
}

@ARTICLE{Kounkel2019AJ,
       author = {{Kounkel}, Marina and {Covey}, Kevin and {Moe}, Maxwell and {Kratter}, Kaitlin M. and {Su{\'a}rez}, Genaro and {Stassun}, Keivan G. and {Rom{\'a}n-Z{\'u}{\~n}iga}, Carlos and {Hernandez}, Jesus and {Kim}, Jinyoung Serena and {Pe{\~n}a Ram{\'\i}rez}, Karla and {Roman-Lopes}, Alexandre and {Stringfellow}, Guy S. and {Jaehnig}, Karl O. and {Borissova}, Jura and {Tofflemire}, Benjamin and {Krolikowski}, Daniel and {Rizzuto}, Aaron and {Kraus}, Adam and {Badenes}, Carles and {Longa-Pe{\~n}a}, Pen{\'e}lope and {G{\'o}mez Maqueo Chew}, Yilen and {Barba}, Rodolfo and {Nidever}, David L. and {Brown}, Cody and {De Lee}, Nathan and {Pan}, Kaike and {Bizyaev}, Dmitry and {Oravetz}, Daniel and {Oravetz}, Audrey},
        title = "{Close Companions around Young Stars}",
      journal = {\aj},
         year = 2019,
        month = may,
       volume = {157},
       number = {5},
          eid = {196},
        pages = {196},
          doi = {10.3847/1538-3881/ab13b1},
archivePrefix = {arXiv},
       eprint = {1903.10523},
 primaryClass = {astro-ph.SR},
       adsurl = {https://ui.adsabs.harvard.edu/abs/2019AJ....157..196K}
}

@ARTICLE{Patience+2002,
       author = {{Patience}, J. and {Ghez}, A.~M. and {Reid}, I.~N. and {Matthews}, K.},
        title = "{A High Angular Resolution Multiplicity Survey of the Open Clusters {\ensuremath{\alpha}} Persei and Praesepe}",
      journal = {\aj},
         year = 2002,
        month = mar,
       volume = {123},
       number = {3},
        pages = {1570-1602},
          doi = {10.1086/338431},
archivePrefix = {arXiv},
       eprint = {astro-ph/0111156},
 primaryClass = {astro-ph},
       adsurl = {https://ui.adsabs.harvard.edu/abs/2002AJ....123.1570P}
}

@ARTICLE{Torres2021ApJ,
       author = {{Torres}, Guillermo and {Latham}, David W. and {Quinn}, Samuel N.},
        title = "{Long-term Spectroscopic Survey of the Pleiades Cluster: The Binary Population}",
      journal = {\apj},
         year = 2021,
        month = nov,
       volume = {921},
       number = {2},
          eid = {117},
        pages = {117},
          doi = {10.3847/1538-4357/ac1585},
archivePrefix = {arXiv},
       eprint = {2107.10259},
 primaryClass = {astro-ph.SR},
       adsurl = {https://ui.adsabs.harvard.edu/abs/2021ApJ...921..117T}
}

@ARTICLE{Cohen2020AJ,
       author = {{Cohen}, Roger E. and {Geller}, Aaron M. and {von Hippel}, Ted},
        title = "{Bayesian Characterization of Main-sequence Binaries in the Old Open Cluster NGC 188}",
      journal = {\aj},
         year = 2020,
        month = jan,
       volume = {159},
       number = {1},
          eid = {11},
        pages = {11},
          doi = {10.3847/1538-3881/ab59d7},
archivePrefix = {arXiv},
       eprint = {1911.08390},
 primaryClass = {astro-ph.SR},
       adsurl = {https://ui.adsabs.harvard.edu/abs/2020AJ....159...11C}
}

@ARTICLE{Geller2021AJ,
       author = {{Geller}, Aaron M. and {Mathieu}, Robert D. and {Latham}, David W. and {Pollack}, Maxwell and {Torres}, Guillermo and {Leiner}, Emily M.},
        title = "{Stellar Radial Velocities in the Old Open Cluster M67 (NGC 2682). II. The Spectroscopic Binary Population}",
      journal = {\aj},
         year = 2021,
        month = apr,
       volume = {161},
       number = {4},
          eid = {190},
        pages = {190},
          doi = {10.3847/1538-3881/abdd23},
archivePrefix = {arXiv},
       eprint = {2101.07883},
 primaryClass = {astro-ph.SR},
       adsurl = {https://ui.adsabs.harvard.edu/abs/2021AJ....161..190G}
}

@ARTICLE{Kroupa1993MNRAS,
       author = {{Kroupa}, Pavel and {Tout}, Christopher A. and {Gilmore}, Gerard},
        title = "{The Distribution of Low-Mass Stars in the Galactic Disc}",
      journal = {\mnras},
         year = 1993,
        month = jun,
       volume = {262},
        pages = {545-587},
          doi = {10.1093/mnras/262.3.545},
       adsurl = {https://ui.adsabs.harvard.edu/abs/1993MNRAS.262..545K}
}

@ARTICLE{Bate2012MNRAS,
       author = {{Bate}, Matthew R.},
        title = "{Stellar, brown dwarf and multiple star properties from a radiation hydrodynamical simulation of star cluster formation}",
      journal = {\mnras},
         year = 2012,
        month = feb,
       volume = {419},
       number = {4},
        pages = {3115-3146},
          doi = {10.1111/j.1365-2966.2011.19955.x},
archivePrefix = {arXiv},
       eprint = {1110.1092},
 primaryClass = {astro-ph.SR},
       adsurl = {https://ui.adsabs.harvard.edu/abs/2012MNRAS.419.3115B}
}

@MISC{gaiaedr3_doc,
       author = {{van Leeuwen}, F. and {de Bruijne}, J. and {Babusiaux}, C. and {Casta{\~n}eda}, J. and {Hobbs}, D. and {Busso}, G. and {Sartoretti}, P. and {Utrilla}, E. and {Luri}, X. and {Marrese}, P.~M. and {Mora}, A. and {Fabricius}, C. and {Gonz{\'a}lez-N{\'u}{\~n}ez}, J. and {Hambly}, N. and {Altavilla}, G. and {Altmann}, M. and {Antoja}, T. and {Arenou}, F. and {Bakker}, J. and {Balbinot}, E. and {Barache}, C. and {Bastian}, U. and {Bauchet}, N. and {Bellazzini}, M. and {Biermann}, M. and {Blomme}, R. and {Bombrun}, A. and {Brown}, A. and {Busonero}, D. and {Butkevich}, A. and {Cacciari}, C. and {Carrasco}, J.~M. and {Cheek}, N. and {Clotet}, M. and {Creevey}, O. and {Crowley}, C. and {C{\'a}novas}, H. and {David}, M. and {Davidson}, M. and {De Angeli}, F. and {Diakit{\'e}}, S. and {Drimmel}, R. and {Duran}, J. and {Evans}, D.~W. and {Fabrizio}, M. and {Fern{\'a}ndez-Hern{\'a}ndez}, J. and {Figueras}, F. and {Findeisen}, K. and {Garcia-Gutierrez}, A. and {Gracia-Abril. G.} and {Guerra}, R. and {Guti{\'e}rrez-S{\'a}nchez}, R. and {Helmi}, A. and {Henar Sarmiento}, M. and {Hernandez}, J. and {Hutton}, A. and {Jordi}, C. and {Khanna}, S. and {Klioner}, S. and {Lammers}, U. and {Leclerc}, N. and {Lindegren}, L. and {L{\"o}ffler}, W. and {Marinoni}, S. and {Mart{\'\i}n-Fleitas}, J. and {Masana}, E. and {Masip Vela}, A. and {Masip}, A. and {Messineo}, R. and {Michalik}, D. and {Mignard}, F. and {Montegriffo}, P. and {Muraveva}, T. and {Nienartowicz}, K. and {Pancino}, E. and {Panem}, C. and {Portell}, J. and {Racero}, E. and {Rainer}, M. and {Ramos}, P. and {Reyl{\'e}}, C. and {R{\'\i}os Diaz}, C. and {Riva}, A. and {Robin}, A. and {Robin}, A. and {Roegiers}, T. and {Romero-G{\'o}mez}, M. and {Rowell}, N. and {Rybizki}, J. and {Salgado}, J. and {Sanna}, N. and {Seabroke}, G. and {Segovia}, J.~C. and {Siddiqui}, H. and {Smart}, R. and {Stephenson}, C. and {Teyssier}, D. and {Torra}, F. and {Turon}, C. and {Valero}, J. and {Vallenari}, A. and {van Leeuwen}, M. and {Weiler}, M.},
        title = "{Gaia EDR3 documentation}",
        howpublished = {Gaia EDR3 documentation, European Space Agency; Gaia Data Processing and Analysis Consortium.},
        url={https://gea.esac.esa.int/archive/documentation/GEDR3/index.html},
         year = 2021,
        month = mar,
       adsurl = {https://ui.adsabs.harvard.edu/abs/2021gdr3.reptE....V},
        note = {Accessed: 2024-06-23}
}

@misc{2mass_doc,
        author = "{The 2MASS Team}",
        title = {2MASS All-Sky Data Release Explanatory Supplement: User's Guide},
        year = 2006,
        howpublished = {Infrared Processing and Analysis Center, California Institute of Technology, and University of Massachusetts},
        url = {https://irsa.ipac.caltech.edu/data/2MASS/docs/releases/allsky/doc/sec2_2a.html},
        note = {Accessed: 2024-06-23}
}

@ARTICLE{Skrutskie+2006,
       author = {{Skrutskie}, M.~F. and {Cutri}, R.~M. and {Stiening}, R. and {Weinberg}, M.~D. and {Schneider}, S. and {Carpenter}, J.~M. and {Beichman}, C. and {Capps}, R. and {Chester}, T. and {Elias}, J. and {Huchra}, J. and {Liebert}, J. and {Lonsdale}, C. and {Monet}, D.~G. and {Price}, S. and {Seitzer}, P. and {Jarrett}, T. and {Kirkpatrick}, J.~D. and {Gizis}, J.~E. and {Howard}, E. and {Evans}, T. and {Fowler}, J. and {Fullmer}, L. and {Hurt}, R. and {Light}, R. and {Kopan}, E.~L. and {Marsh}, K.~A. and {McCallon}, H.~L. and {Tam}, R. and {Van Dyk}, S. and {Wheelock}, S.},
        title = "{The Two Micron All Sky Survey (2MASS)}",
      journal = {Astronomical Journal},
         year = 2006,
        month = feb,
       volume = {131},
       number = {2},
        pages = {1163-1183},
          doi = {10.1086/498708},
       adsurl = {https://ui.adsabs.harvard.edu/abs/2006AJ....131.1163S}
}

@ARTICLE{Wright+2010,
       author = {{Wright}, Edward L. and {Eisenhardt}, Peter R.~M. and {Mainzer}, Amy K. and {Ressler}, Michael E. and {Cutri}, Roc M. and {Jarrett}, Thomas and {Kirkpatrick}, J. Davy and {Padgett}, Deborah and {McMillan}, Robert S. and {Skrutskie}, Michael and {Stanford}, S.~A. and {Cohen}, Martin and {Walker}, Russell G. and {Mather}, John C. and {Leisawitz}, David and {Gautier}, Thomas N., III and {McLean}, Ian and {Benford}, Dominic and {Lonsdale}, Carol J. and {Blain}, Andrew and {Mendez}, Bryan and {Irace}, William R. and {Duval}, Valerie and {Liu}, Fengchuan and {Royer}, Don and {Heinrichsen}, Ingolf and {Howard}, Joan and {Shannon}, Mark and {Kendall}, Martha and {Walsh}, Amy L. and {Larsen}, Mark and {Cardon}, Joel G. and {Schick}, Scott and {Schwalm}, Mark and {Abid}, Mohamed and {Fabinsky}, Beth and {Naes}, Larry and {Tsai}, Chao-Wei},
        title = "{The Wide-field Infrared Survey Explorer (WISE): Mission Description and Initial On-orbit Performance}",
      journal = {Astronomical Journal},
         year = 2010,
        month = dec,
       volume = {140},
       number = {6},
        pages = {1868-1881},
          doi = {10.1088/0004-6256/140/6/1868},
archivePrefix = {arXiv},
       eprint = {1008.0031},
 primaryClass = {astro-ph.IM},
       adsurl = {https://ui.adsabs.harvard.edu/abs/2010AJ....140.1868W}
}

@ARTICLE{Mainzer2011ApJ,
       author = {{Mainzer}, A. and {Bauer}, J. and {Grav}, T. and {Masiero}, J. and {Cutri}, R.~M. and {Dailey}, J. and {Eisenhardt}, P. and {McMillan}, R.~S. and {Wright}, E. and {Walker}, R. and {Jedicke}, R. and {Spahr}, T. and {Tholen}, D. and {Alles}, R. and {Beck}, R. and {Brandenburg}, H. and {Conrow}, T. and {Evans}, T. and {Fowler}, J. and {Jarrett}, T. and {Marsh}, K. and {Masci}, F. and {McCallon}, H. and {Wheelock}, S. and {Wittman}, M. and {Wyatt}, P. and {DeBaun}, E. and {Elliott}, G. and {Elsbury}, D. and {Gautier}, IV, T. and {Gomillion}, S. and {Leisawitz}, D. and {Maleszewski}, C. and {Micheli}, M. and {Wilkins}, A.},
        title = "{Preliminary Results from NEOWISE: An Enhancement to the Wide-field Infrared Survey Explorer for Solar System Science}",
      journal = {\apj},
         year = 2011,
        month = apr,
       volume = {731},
       number = {1},
          eid = {53},
        pages = {53},
          doi = {10.1088/0004-637X/731/1/53},
archivePrefix = {arXiv},
       eprint = {1102.1996},
 primaryClass = {astro-ph.EP},
       adsurl = {https://ui.adsabs.harvard.edu/abs/2011ApJ...731...53M}
}

@article{astropy:2013,
        Adsurl = {http://adsabs.harvard.edu/abs/2013A%26A...558A..33A},
        Archiveprefix = {arXiv},
        Author = {{Astropy Collaboration} and {Robitaille}, T.~P. and {Tollerud}, E.~J. and {Greenfield}, P. and {Droettboom}, M. and {Bray}, E. and {Aldcroft}, T. and {Davis}, M. and {Ginsburg}, A. and {Price-Whelan}, A.~M. and {Kerzendorf}, W.~E. and {Conley}, A. and {Crighton}, N. and {Barbary}, K. and {Muna}, D. and {Ferguson}, H. and {Grollier}, F. and {Parikh}, M.~M. and {Nair}, P.~H. and {Unther}, H.~M. and {Deil}, C. and {Woillez}, J. and {Conseil}, S. and {Kramer}, R. and {Turner}, J.~E.~H. and {Singer}, L. and {Fox}, R. and {Weaver}, B.~A. and {Zabalza}, V. and {Edwards}, Z.~I. and {Azalee Bostroem}, K. and {Burke}, D.~J. and {Casey}, A.~R. and {Crawford}, S.~M. and {Dencheva}, N. and {Ely}, J. and {Jenness}, T. and {Labrie}, K. and {Lim}, P.~L. and {Pierfederici}, F. and {Pontzen}, A. and {Ptak}, A. and {Refsdal}, B. and {Servillat}, M. and {Streicher}, O.},
        Doi = {10.1051/0004-6361/201322068},
        Eid = {A33},
        Eprint = {1307.6212},
        Journal = {\aap},
        Month = oct,
        Pages = {A33},
        Primaryclass = {astro-ph.IM},
        Title = {{Astropy: A community Python package for astronomy}},
        Volume = 558,
        Year = 2013
}

@article{astropy:2018,
        Adsurl = {https://ui.adsabs.harvard.edu/#abs/2018AJ....156..123T},
        Author = {{Price-Whelan}, A.~M. and {Sip{\H{o}}cz}, B.~M. and {G{\"u}nther}, H.~M. and {Lim}, P.~L. and {Crawford}, S.~M. and {Conseil}, S. and {Shupe}, D.~L. and {Craig}, M.~W. and {Dencheva}, N. and {Ginsburg}, A. and {VanderPlas}, J.~T. and {Bradley}, L.~D. and {P{\'e}rez-Su{\'a}rez}, D. and {de Val-Borro}, M. and {Paper Contributors}, (Primary and {Aldcroft}, T.~L. and {Cruz}, K.~L. and {Robitaille}, T.~P. and {Tollerud}, E.~J. and {Coordination Committee}, (Astropy and {Ardelean}, C. and {Babej}, T. and {Bach}, Y.~P. and {Bachetti}, M. and {Bakanov}, A.~V. and {Bamford}, S.~P. and {Barentsen}, G. and {Barmby}, P. and {Baumbach}, A. and {Berry}, K.~L. and {Biscani}, F. and {Boquien}, M. and {Bostroem}, K.~A. and {Bouma}, L.~G. and {Brammer}, G.~B. and {Bray}, E.~M. and {Breytenbach}, H. and {Buddelmeijer}, H. and {Burke}, D.~J. and {Calderone}, G. and {Cano Rodr{\'\i}guez}, J.~L. and {Cara}, M. and {Cardoso}, J.~V.~M. and {Cheedella}, S. and {Copin}, Y. and {Corrales}, L. and {Crichton}, D. and {D{\textquoteright}Avella}, D. and {Deil}, C. and {Depagne}, {\'E}. and {Dietrich}, J.~P. and {Donath}, A. and {Droettboom}, M. and {Earl}, N. and {Erben}, T. and {Fabbro}, S. and {Ferreira}, L.~A. and {Finethy}, T. and {Fox}, R.~T. and {Garrison}, L.~H. and {Gibbons}, S.~L.~J. and {Goldstein}, D.~A. and {Gommers}, R. and {Greco}, J.~P. and {Greenfield}, P. and {Groener}, A.~M. and {Grollier}, F. and {Hagen}, A. and {Hirst}, P. and {Homeier}, D. and {Horton}, A.~J. and {Hosseinzadeh}, G. and {Hu}, L. and {Hunkeler}, J.~S. and {Ivezi{\'c}}, {\v{Z}}. and {Jain}, A. and {Jenness}, T. and {Kanarek}, G. and {Kendrew}, S. and {Kern}, N.~S. and {Kerzendorf}, W.~E. and {Khvalko}, A. and {King}, J. and {Kirkby}, D. and {Kulkarni}, A.~M. and {Kumar}, A. and {Lee}, A. and {Lenz}, D. and {Littlefair}, S.~P. and {Ma}, Z. and {Macleod}, D.~M. and {Mastropietro}, M. and {McCully}, C. and {Montagnac}, S. and {Morris}, B.~M. and {Mueller}, M. and {Mumford}, S.~J. and {Muna}, D. and {Murphy}, N.~A. and {Nelson}, S. and {Nguyen}, G.~H. and {Ninan}, J.~P. and {N{\"o}the}, M. and {Ogaz}, S. and {Oh}, S. and {Parejko}, J.~K. and {Parley}, N. and {Pascual}, S. and {Patil}, R. and {Patil}, A.~A. and {Plunkett}, A.~L. and {Prochaska}, J.~X. and {Rastogi}, T. and {Reddy Janga}, V. and {Sabater}, J. and {Sakurikar}, P. and {Seifert}, M. and {Sherbert}, L.~E. and {Sherwood-Taylor}, H. and {Shih}, A.~Y. and {Sick}, J. and {Silbiger}, M.~T. and {Singanamalla}, S. and {Singer}, L.~P. and {Sladen}, P.~H. and {Sooley}, K.~A. and {Sornarajah}, S. and {Streicher}, O. and {Teuben}, P. and {Thomas}, S.~W. and {Tremblay}, G.~R. and {Turner}, J.~E.~H. and {Terr{\'o}n}, V. and {van Kerkwijk}, M.~H. and {de la Vega}, A. and {Watkins}, L.~L. and {Weaver}, B.~A. and {Whitmore}, J.~B. and {Woillez}, J. and {Zabalza}, V. and {Contributors}, (Astropy},
        Doi = {10.3847/1538-3881/aabc4f},
        Eid = {123},
        Journal = {\aj},
        Month = Sep,
        Pages = {123},
        Primaryclass = {astro-ph.IM},
        Title = {{The Astropy Project: Building an Open-science Project and Status of the v2.0 Core Package}},
        Volume = {156},
        Year = 2018
}

@ARTICLE{astropy:2022,
       author = {{Astropy Collaboration} and {Price-Whelan}, Adrian M. and {Lim}, Pey Lian and {Earl}, Nicholas and {Starkman}, Nathaniel and {Bradley}, Larry and {Shupe}, David L. and {Patil}, Aarya A. and {Corrales}, Lia and {Brasseur}, C.~E. and {N{\"o}the}, Maximilian and {Donath}, Axel and {Tollerud}, Erik and {Morris}, Brett M. and {Ginsburg}, Adam and {Vaher}, Eero and {Weaver}, Benjamin A. and {Tocknell}, James and {Jamieson}, William and {van Kerkwijk}, Marten H. and {Robitaille}, Thomas P. and {Merry}, Bruce and {Bachetti}, Matteo and {G{\"u}nther}, H. Moritz and {Aldcroft}, Thomas L. and {Alvarado-Montes}, Jaime A. and {Archibald}, Anne M. and {B{\'o}di}, Attila and {Bapat}, Shreyas and {Barentsen}, Geert and {Baz{\'a}n}, Juanjo and {Biswas}, Manish and {Boquien}, M{\'e}d{\'e}ric and {Burke}, D.~J. and {Cara}, Daria and {Cara}, Mihai and {Conroy}, Kyle E. and {Conseil}, Simon and {Craig}, Matthew W. and {Cross}, Robert M. and {Cruz}, Kelle L. and {D'Eugenio}, Francesco and {Dencheva}, Nadia and {Devillepoix}, Hadrien A.~R. and {Dietrich}, J{\"o}rg P. and {Eigenbrot}, Arthur Davis and {Erben}, Thomas and {Ferreira}, Leonardo and {Foreman-Mackey}, Daniel and {Fox}, Ryan and {Freij}, Nabil and {Garg}, Suyog and {Geda}, Robel and {Glattly}, Lauren and {Gondhalekar}, Yash and {Gordon}, Karl D. and {Grant}, David and {Greenfield}, Perry and {Groener}, Austen M. and {Guest}, Steve and {Gurovich}, Sebastian and {Handberg}, Rasmus and {Hart}, Akeem and {Hatfield-Dodds}, Zac and {Homeier}, Derek and {Hosseinzadeh}, Griffin and {Jenness}, Tim and {Jones}, Craig K. and {Joseph}, Prajwel and {Kalmbach}, J. Bryce and {Karamehmetoglu}, Emir and {Ka{\l}uszy{\'n}ski}, Miko{\l}aj and {Kelley}, Michael S.~P. and {Kern}, Nicholas and {Kerzendorf}, Wolfgang E. and {Koch}, Eric W. and {Kulumani}, Shankar and {Lee}, Antony and {Ly}, Chun and {Ma}, Zhiyuan and {MacBride}, Conor and {Maljaars}, Jakob M. and {Muna}, Demitri and {Murphy}, N.~A. and {Norman}, Henrik and {O'Steen}, Richard and {Oman}, Kyle A. and {Pacifici}, Camilla and {Pascual}, Sergio and {Pascual-Granado}, J. and {Patil}, Rohit R. and {Perren}, Gabriel I. and {Pickering}, Timothy E. and {Rastogi}, Tanuj and {Roulston}, Benjamin R. and {Ryan}, Daniel F. and {Rykoff}, Eli S. and {Sabater}, Jose and {Sakurikar}, Parikshit and {Salgado}, Jes{\'u}s and {Sanghi}, Aniket and {Saunders}, Nicholas and {Savchenko}, Volodymyr and {Schwardt}, Ludwig and {Seifert-Eckert}, Michael and {Shih}, Albert Y. and {Jain}, Anany Shrey and {Shukla}, Gyanendra and {Sick}, Jonathan and {Simpson}, Chris and {Singanamalla}, Sudheesh and {Singer}, Leo P. and {Singhal}, Jaladh and {Sinha}, Manodeep and {Sip{\H{o}}cz}, Brigitta M. and {Spitler}, Lee R. and {Stansby}, David and {Streicher}, Ole and {{\v{S}}umak}, Jani and {Swinbank}, John D. and {Taranu}, Dan S. and {Tewary}, Nikita and {Tremblay}, Grant R. and {de Val-Borro}, Miguel and {Van Kooten}, Samuel J. and {Vasovi{\'c}}, Zlatan and {Verma}, Shresth and {de Miranda Cardoso}, Jos{\'e} Vin{\'\i}cius and {Williams}, Peter K.~G. and {Wilson}, Tom J. and {Winkel}, Benjamin and {Wood-Vasey}, W.~M. and {Xue}, Rui and {Yoachim}, Peter and {Zhang}, Chen and {Zonca}, Andrea and {Astropy Project Contributors}},
        title = "{The Astropy Project: Sustaining and Growing a Community-oriented Open-source Project and the Latest Major Release (v5.0) of the Core Package}",
      journal = {\apj},
         year = 2022,
        month = aug,
       volume = {935},
       number = {2},
          eid = {167},
        pages = {167},
          doi = {10.3847/1538-4357/ac7c74},
archivePrefix = {arXiv},
       eprint = {2206.14220},
 primaryClass = {astro-ph.IM},
       adsurl = {https://ui.adsabs.harvard.edu/abs/2022ApJ...935..167A}
}

@ARTICLE{GaiaDR3,
       author = {{Gaia Collaboration} and {Vallenari}, A. and {Brown}, A.~G.~A. and {Prusti}, T. and {de Bruijne}, J.~H.~J. and {Arenou}, F. and {Babusiaux}, C. and {Biermann}, M. and {Creevey}, O.~L. and {Ducourant}, C. and {Evans}, D.~W. and {Eyer}, L. and {Guerra}, R. and {Hutton}, A. and {Jordi}, C. and {Klioner}, S.~A. and {Lammers}, U.~L. and {Lindegren}, L. and {Luri}, X. and {Mignard}, F. and {Panem}, C. and {Pourbaix}, D. and {Randich}, S. and {Sartoretti}, P. and {Soubiran}, C. and {Tanga}, P. and {Walton}, N.~A. and {Bailer-Jones}, C.~A.~L. and {Bastian}, U. and {Drimmel}, R. and {Jansen}, F. and {Katz}, D. and {Lattanzi}, M.~G. and {van Leeuwen}, F. and {Bakker}, J. and {Cacciari}, C. and {Casta{\~n}eda}, J. and {De Angeli}, F. and {Fabricius}, C. and {Fouesneau}, M. and {Fr{\'e}mat}, Y. and {Galluccio}, L. and {Guerrier}, A. and {Heiter}, U. and {Masana}, E. and {Messineo}, R. and {Mowlavi}, N. and {Nicolas}, C. and {Nienartowicz}, K. and {Pailler}, F. and {Panuzzo}, P. and {Riclet}, F. and {Roux}, W. and {Seabroke}, G.~M. and {Sordo}, R. and {Th{\'e}venin}, F. and {Gracia-Abril}, G. and {Portell}, J. and {Teyssier}, D. and {Altmann}, M. and {Andrae}, R. and {Audard}, M. and {Bellas-Velidis}, I. and {Benson}, K. and {Berthier}, J. and {Blomme}, R. and {Burgess}, P.~W. and {Busonero}, D. and {Busso}, G. and {C{\'a}novas}, H. and {Carry}, B. and {Cellino}, A. and {Cheek}, N. and {Clementini}, G. and {Damerdji}, Y. and {Davidson}, M. and {de Teodoro}, P. and {Nu{\~n}ez Campos}, M. and {Delchambre}, L. and {Dell'Oro}, A. and {Esquej}, P. and {Fern{\'a}ndez-Hern{\'a}ndez}, J. and {Fraile}, E. and {Garabato}, D. and {Garc{\'\i}a-Lario}, P. and {Gosset}, E. and {Haigron}, R. and {Halbwachs}, J. -L. and {Hambly}, N.~C. and {Harrison}, D.~L. and {Hern{\'a}ndez}, J. and {Hestroffer}, D. and {Hodgkin}, S.~T. and {Holl}, B. and {Jan{\ss}en}, K. and {Jevardat de Fombelle}, G. and {Jordan}, S. and {Krone-Martins}, A. and {Lanzafame}, A.~C. and {L{\"o}ffler}, W. and {Marchal}, O. and {Marrese}, P.~M. and {Moitinho}, A. and {Muinonen}, K. and {Osborne}, P. and {Pancino}, E. and {Pauwels}, T. and {Recio-Blanco}, A. and {Reyl{\'e}}, C. and {Riello}, M. and {Rimoldini}, L. and {Roegiers}, T. and {Rybizki}, J. and {Sarro}, L.~M. and {Siopis}, C. and {Smith}, M. and {Sozzetti}, A. and {Utrilla}, E. and {van Leeuwen}, M. and {Abbas}, U. and {{\'A}brah{\'a}m}, P. and {Abreu Aramburu}, A. and {Aerts}, C. and {Aguado}, J.~J. and {Ajaj}, M. and {Aldea-Montero}, F. and {Altavilla}, G. and {{\'A}lvarez}, M.~A. and {Alves}, J. and {Anders}, F. and {Anderson}, R.~I. and {Anglada Varela}, E. and {Antoja}, T. and {Baines}, D. and {Baker}, S.~G. and {Balaguer-N{\'u}{\~n}ez}, L. and {Balbinot}, E. and {Balog}, Z. and {Barache}, C. and {Barbato}, D. and {Barros}, M. and {Barstow}, M.~A. and {Bartolom{\'e}}, S. and {Bassilana}, J. -L. and {Bauchet}, N. and {Becciani}, U. and {Bellazzini}, M. and {Berihuete}, A. and {Bernet}, M. and {Bertone}, S. and {Bianchi}, L. and {Binnenfeld}, A. and {Blanco-Cuaresma}, S. and {Blazere}, A. and {Boch}, T. and {Bombrun}, A. and {Bossini}, D. and {Bouquillon}, S. and {Bragaglia}, A. and {Bramante}, L. and {Breedt}, E. and {Bressan}, A. and {Brouillet}, N. and {Brugaletta}, E. and {Bucciarelli}, B. and {Burlacu}, A. and {Butkevich}, A.~G. and {Buzzi}, R. and {Caffau}, E. and {Cancelliere}, R. and {Cantat-Gaudin}, T. and {Carballo}, R. and {Carlucci}, T. and {Carnerero}, M.~I. and {Carrasco}, J.~M. and {Casamiquela}, L. and {Castellani}, M. and {Castro-Ginard}, A. and {Chaoul}, L. and {Charlot}, P. and {Chemin}, L. and {Chiaramida}, V. and {Chiavassa}, A. and {Chornay}, N. and {Comoretto}, G. and {Contursi}, G. and {Cooper}, W.~J. and {Cornez}, T. and {Cowell}, S. and {Crifo}, F. and {Cropper}, M. and {Crosta}, M. and {Crowley}, C. and {Dafonte}, C. and {Dapergolas}, A. and {David}, M. and {David}, P. and {de Laverny}, P. and {De Luise}, F. and {De March}, R.},
        title = "{Gaia Data Release 3. Summary of the content and survey properties}",
      journal = {\aap},
         year = 2023,
        month = jun,
       volume = {674},
          eid = {A1},
        pages = {A1},
          doi = {10.1051/0004-6361/202243940},
archivePrefix = {arXiv},
       eprint = {2208.00211},
 primaryClass = {astro-ph.GA},
       adsurl = {https://ui.adsabs.harvard.edu/abs/2023A&A...674A...1G}
}
\bibliographystyle{aasjournal}


\listofchanges

\end{document}